\documentclass[12pt,titlepage]{utarticle}
\usepackage{amssymb,graphicx}
\usepackage{amsmath}
\usepackage[margin=0.5cm]{caption}
\usepackage{hyperref} 
\usepackage{cite}
\usepackage{upgreek}

\def\e{{\rm e}}

\def\eps{\epsilon}
\def\d{\partial}
\def\l{\left(}
\def\r{\right)}

\newcommand{\be}{\begin{equation}}
\newcommand{\ee}{\end{equation}}
\newcommand{\bea}{\begin{eqnarray}}
\newcommand{\eea}{\end{eqnarray}}
\newcommand{\bg}{\begin{gather}}
\newcommand{\eg}{\end{gather}}
\newcommand{\bseq}{\begin{subequations}}
\newcommand{\eseq}{\end{subequations}}

\renewcommand{\ln}{\mathop{\rm ln}\nolimits}

\newcommand{\Tr}{{\rm Tr}}
\def\half{\frac{1}{2}}

\newcommand{\bra}[1]{\langle #1 |}
\newcommand{\ket}[1]{| #1 \rangle}

\begin{document}
\baselineskip=15.5pt
\title{Flux Tube Spectra
  from Approximate\\[0.3cm] Integrability at Low Energies\footnote{Prepared for a special issue of JETP dedicated to the 60th birthday of Valery Rubakov.} }

\author{Sergei Dubovsky
   \address{
      Center for Cosmology and Particle Physics, \\Department of Physics,\\
      New York University\\ 
      New York, NY, 10003, USA\\
   }
   , Raphael Flauger $^{1,}$ 
   \address{
      School of Natural Sciences,\\ 
      Institute for Advanced Study,\\
      Princeton, NJ 08540, USA\\
      {~}\\[.5cm]
      {\rm {\it e}-mail}\\[.1cm]
      \emailt{dubovsky@nyu.edu}\\
      \emailt{flauger@nyu.edu}\\
      \emailt{gorbenko@nyu.edu\\
    %  ~\\
      ~~~~~~~~~~~~~~~~~~~~~~~~~~~{\large\it To Valery Rubakov\\
~~~~~~~~~~~~~~~~~~~~~~~~~~~~~~~~~~with gratitude      }}\\
   } and Victor Gorbenko $^1$
}
%{ \large
% Sergei~Dubovsky$^a$, Raphael Flauger$^{a,b}$, and Victor Gorbenko$^a$}\\
%\vspace{.2cm}
%{\small  \textit{$^a$Center for Cosmology and Particle Physics, Department of Physics, \\New York University, New York, NY, 10003, USA }}\\
%{\small  \textit{$^b$ School of Natural Sciences, Institute for Advanced Study, Princeton, NJ 08540, USA}}

%\end{center}
%\begin{center}

\Abstract{We provide a detailed introduction to a method we recently proposed for calculating the spectrum of excitations of effective strings such as QCD flux tubes.
The method relies  on the approximate integrability of the low energy effective theory describing the flux tube excitations and is is based on the Thermodynamic Bethe Ansatz (TBA). The approximate integrability is a consequence of the Lorentz symmetry of QCD.
For excited states the convergence of the TBA technique is significantly better than that of the traditional perturbative approach.
We apply the new technique to the lattice spectra for fundamental flux tubes in gluodynamics in D=3+1 and D=2+1, and to $k$-strings in gluodynamics in D=2+1. We identify a massive pseudoscalar resonance on the world sheet of the confining strings in $SU(3)$ gluodynamics in D=3+1, and massive scalar resonances on the world sheet of $k=2,3$ strings in $SU(6)$ gluodynamics in D=2+1.\vskip3mm
}

\maketitle
\tableofcontents
\section{Introduction}
String theory originated as a candidate theory of strong interactions \cite{Veneziano:1968yb}.  
However, it was soon abandoned as a theory of hadrons, at least for the time being, because it failed to reproduce the observed properties of deep inelastic scattering as well as the asymptotic freedom of non-Abelian gauge theories. 
Regardless, the success of the Veneziano amplitude in describing many aspects of the hadron spectrum and scattering is hardly a coincidence. Confining strings (flux tubes) are crucial ingredients in the strongly coupled QCD dynamics responsible for color confinement, and their presence is vividly revealed by lattice QCD simulations ~\cite{Bissey:2006bz}\footnote{See 
\url{http://www.physics.adelaide.edu.au/theory/staff/leinweber/VisualQCD/Nobel/} for animations.}, suggesting that understanding the structure and dynamics of QCD flux tubes might provide insights into the dynamics of color confinement.

%It remains unclear whether QCD allows a dual reformulation with strings as fundamental degrees of freedom. Indeed, the mere existence of string-like excitations in a model does not imply that it allows a useful {\it microscopic} string theory description. Abrikosov--Nielsen--Olesen vortices in the Abelian Higgs model provide an example where such a description is highly unlikely. However, independent of the existence of a reformulation, understanding the structure of the low energy theory localized on the world sheet of the QCD flux tube is a well posed and interesting question in its own right and answering it may provide new insights into the dynamics of color confinement.

 The modern approach to the relation between string theory and gauge theories relies on the AdS/CFT correspondence \cite{Maldacena:1997re}. Within this framework the QCD flux tube is expected to be described by a string propagating in a space-time with an extra curved dimension, which can be interpreted as the dynamical string tension, or equivalently, the renormalization group scale \cite{Polyakov:1998ju}. Identifying a concrete string theory which would provide a holographic description of non-supersymmetric QCD remains  a long shot, and even if this dual string theory were found, it would be outside the regime in which we currently have theoretical control. 
%Also, describing the asymptotically free regime in this language, if possible at all, remains a major challenge.

In this paper we thus focus on a rather direct path towards understanding the structure of the flux tube theory that does not involve holography. Instead, it is based on existing lattice techniques combined with effective field theory and tools from integrability. 

Advances in lattice QCD simulations have allowed to measure the spectrum of low lying world sheet excitations with impressive accuracy \cite{Athenodorou:2010cs,Athenodorou:2011rx,Athenodorou:2013ioa}. However, until now the theoretical interpretation of these results was problematic. For most states the string lengths accessible in the lattice simulations were too short for existing techniques to be reliable. The conventional perturbative methods  \cite{Luscher:1980ac,Luscher:2004ib,Aharony:2010db} for calculating the spectrum of string excitations result in badly diverging asymptotic series in this regime, preventing the interpretation of the data.
At the same time, the data exhibited a number of puzzling and suggestive features. In particular, while perturbative calculations were not reliable, many of the levels show surprisingly good agreement with the spectrum of a free bosonic string quantized in the light cone gauge following the classic paper \cite{Goddard:1973qh} by Goddard, Goldstone, Rebbi and Thorn  (GGRT) (see also \cite{Arvis:1983fp}). This is confusing, given that the GGRT spectrum is well known to be incompatible with the bulk Poincar{\'e} symmetry if the number of space-time dimensions is different from 26.
 
For the lattice simulations, the computational cost grows exponentially with the length of the string. At least with the current technology this makes it essentially impossible to push lattice calculations into the regime in which conventional perturbation techniques converge. Alternative techniques for calculating the flux tube spectra are thus required which provide better convergence for relatively short strings. We proposed such a technique in~\cite{Dubovsky:2013gi}, and its success relies on the observation that the world sheet theory becomes integrable at low energies. This technique seems sufficient to explain the previously puzzling features seen in the lattice results. In addition, it allowed to show that existing lattice data provide strong evidence for the existence of a massive pseudoscalar state on the world sheet of the QCD flux tube, the world sheet axion.

The goal of the present paper is to provide a detailed account of the method proposed in~\cite{Dubovsky:2013gi}. In section~\ref{GGRT}, we begin with a brief summary of the lattice results and of the effective string theory approach (for a detailed recent review see \cite{Aharony:2013ipa}). We review the results of the conventional perturbative expansion for energy levels, which exhibits a large number of  universal terms.
We explain that the GGRT spectrum, in spite of being inconsistent with the bulk Poincar{\'e} symmetry, still represents a finite volume spectrum of a certain integrable relativistic {\it two-dimensional} theory.  
As we explain, this observation immediately allows to calculate all the universal terms in the spectrum of relativistic effective strings \cite{Dubovsky:2012sh}.
 
 In section \ref{sec:diagrams} we present the new method for calculating the flux tube spectrum. 
 The main idea of the method is to divide the calculation into two steps. First, one perturbatively calculates the
 world sheet $S$-matrix describing the scattering of the flux tube excitations within the effective string theory. One then determines the corresponding finite volume spectrum using the excited state thermodynamic Bethe Ansatz (TBA) \cite{Zamolodchikov:1989cf,Dorey:1996re}, which is very similar to the techniques developed by L\"uscher \cite{Luscher:1985dn,Luscher:1986pf} which are routinely used to extract four-dimensional scattering amplitudes from the lattice QCD data. We provide a partial diagrammatic interpretation of the perturbative resummation performed by the TBA and explain why it is natural to expect that this method results in a better behaved perturbation theory for excited states.
 
 In section \ref{sec:levels} we use this technique to interpret the lattice data. We provide more details how to implement the method than in~\cite{Dubovsky:2013gi} and include a larger set of excited states in our analysis. This extended analysis confirms the conclusion reached in \cite{Dubovsky:2013gi}: the lattice data provides strong evidence for the existence of a  pseudoscalar state bound to a confining string. We also apply the technique to the available data for three-dimensional gluodynamics. There we find no evidence for any massive excitations on the fundamental flux tube, but identify massive scalar excitations on $k$-strings.
 
 We conclude in section \ref{sec:last} by discussing future directions and prospects. We also present an intriguing hint for the existence of additional light bound states, coming from the precision ground state data.
  
\section{Lattice Data versus Conventional Perturbative Expansion}
\label{GGRT}

Let us start with a brief summary of lattice results for the excitation spectrum of confining flux tubes. A detailed description of these results and techniques can be found in \cite{Athenodorou:2010cs,Athenodorou:2013ioa,Athenodorou:2011rx}, for a review, see~\cite{Lucini:2012gg}. Most of our discussion will assume that the space-time dimension is $D=4$. However, we will also apply our techniques to the available $D=3$ data. We are interested in the internal dynamics of a single closed flux tube, rather than in effects coming from its boundaries and from interactions between several flux tubes. To discuss these separately, it is necessary to suppress processes where the flux tube can break. This is achieved by performing simulations in pure gluodynamics without dynamical quarks.
Gauge invariant operators in a pure glue theory are constructed as traces of path-ordered exponents of the gauge field $A_\mu$ (Wilson loops),
\be
\label{Op}
{\cal O}_P=\Tr\,P\left(\exp{\int_C A}\right)\;,
\ee
where $C$ is a closed path. In what follows we mostly discuss flux tubes carrying a single unit of fundamental flux. This amounts to  taking the trace in (\ref{Op}) in the fundamental representation of the gauge group.

A nice trick, which allows to concentrate on the dynamics of long flux tubes, is to make use of the non-trivial lattice topology. Namely, one considers states created by operators of the form (\ref{Op}), such that the corresponding path  winds around one of the lattice dimensions. It is convenient to think about the corresponding direction as a spatial, though, of course, on the lattice all directions are Euclidean anyway. States of this kind are orthogonal to conventional glueball states, created by operators (\ref{Op}) with contractible paths.
This follows from a global  $Z_{N}$ symmetry (center symmetry) present in the $SU(N)$ Yang--Mills theory compactified on a circle. It is generated by gauge transformations such that the corresponding gauge functions satisfy twisted boundary conditions. The twist is performed using a multiplication by an element from the center of the gauge group,
\be
\label{twist}
g(R)=e^{2\pi k i/N}g(0)\;,
\ee
where $k$ is an integer.

Transformations satisfying the boundary condition (\ref{twist}) act properly on the gauge configurations and preserve the action functional, but do not originate from a well-defined gauge function.
Hence they should be considered as generating a global, rather than gauge, symmetry. 
Any two transformations with the same twist $k$ are equivalent up to a conventional gauge transformations, hence the resulting symmetry group is  $Z_{N}$.
A state created by an operator (\ref{Op}) with a winding number $k$ carries charge $k$ with respect to this symmetry, so the full Hilbert space splits into a direct sum of $N$ orthogonal subspaces labeled by corresponding winding number (modulo $N$).

Most of the lattice data discussed here is extracted from the two-point correlators of the states carrying a unit charge under the center symmetry (a brief discussion of  $k$-strings with larger values of the charge is presented in section~\ref{sec:kstrings}). These states represent closed flux tubes with a unit winding number around the compact direction. By considering a large enough set of shapes of the Wilson lines one may 
probe not only the ground state but also the low lying excitations of the flux tubes. By measuring the exponential fall-off of the correlators one extracts energies 
of the states created from the vacuum by the corresponding operators, in the same way as for conventional glueball mass measurements.

A theoretical framework for  perturbative calculations of these energies from first principles is provided by effective string theory. The idea is that the flux tube states whose excitation energy above the ground state in the $k=1$ sector is smaller than the mass of the lightest glueball are described by a two-dimensional effective field theory. In the absence of additional symmetries (such as supersymmetry) the only massless degrees of freedom in this theory are Goldstone modes describing the spontaneous breaking of the bulk Poincar{\'e} group $ISO(1,D-1)$ to a residual symmetry group which remains unbroken in the presence of an infinite straight string.
The latter is the product of the world sheet Poincar{\'e} symmetry $ISO(1,1)$ with the transverse rotations $O(D-2)$. This symmetry breaking pattern implies the presence of
$(D-2)$ massless Goldstone degrees of freedom represented by scalar fields $X^i$. Geometrically they parametrize transverse excitations of a flux tube, so that its embedding into the bulk space is given by
\[
X^\mu=(\sigma^\alpha,X^i)\;,
\]
where $\sigma^\alpha$ ($\alpha=1,2$) are the world sheet coordinates. 

The effective action is constructed as  a sum of local geometric invariants corresponding to this embedding, and starts with a Nambu--Goto (NG) term
\begin{gather}
\label{coset_action}
S_{string}=- \ell_s^{-2}\int d^2\sigma\sqrt{-\det{h_{\alpha\beta}}}+\dots=\\ \nonumber= \ell_s^{-2}\int d^2\sigma\l-1-{1\over 2} \d_\alpha X^i\d^\alpha X^i
-{1\over 8}\l\d_\alpha X^i\d^\alpha
X^i\r^2+{1\over 4}\l\d_\alpha X^i\d_\beta X^i\r^2+\dots\r
\end{gather}
where $h_{\alpha\beta}$ is the induced metric on the world sheet,
\be
\label{hind}
h_{\alpha\beta}=\d_\alpha X^\mu\d_\beta X_\mu\;,
\ee
$\ell_s$ is the string scale and $\dots$'s stand for higher order terms.

Within this formalism  the problem of calculating the spectrum of low lying flux tube excitations becomes the computation of the spectrum of low lying Kaluza--Klein (KK) modes of this two-dimensional effective theory upon compactification on a spatial circle of circumference $R$. The traditional approach to this problem is a perturbative expansion in powers of $\ell_s/R$. One perturbatively calculates the spectrum of  a quantum mechanical Hamiltonian obtained after KK
decomposition of the effective action (\ref{coset_action}). At any finite order in the $\ell_s/R$-expansion only a finite number of terms from (\ref{coset_action}) contribute. The procedure is straightforward, even though the algebra may get rather messy as one starts calculating subleading terms in this expansion. The major subtlety in this approach is to enforce  the invariance under non-linearly realized  bulk Lorentz transformations at each order of the expansion,
\be
\label{non_boost}
\updelta^{\alpha i}_\eps X^j=-\epsilon( \delta^{ij}\sigma^\alpha+X^i\d^\alpha X^j)\;,
\ee
where $\epsilon$ is an infinitesimal  parameter of the boost/rotation.
By construction the classical action (\ref{coset_action}) enjoys this symmetry, but depending on the regularization scheme, it may get broken at the intermediate stages of the calculation.

As one can see from (\ref{coset_action}),  a large number of terms in the effective action are fixed as a consequence of non-linearly realized Lorentz transformations (\ref{non_boost}). As a consequence, several leading order terms in the $\ell_s/R$ expansion are universal and can be predicted in a model independent way in any $D$-dimensional theory giving rise to effective string-like objects. The only assumptions entering this prediction is that the bulk theory is relativistic, has a gap, and that the space-time Goldstones $X^i$ are the only massless degrees of freedom carried by the string world sheet. One example of a leading order non-universal term in the effective action (\ref{coset_action}) which does not vanish on-shell\footnote{Or, equivalently, which cannot be removed by a field redefinition.} and is compatible with (\ref{non_boost}) is
\[
\delta S\propto \ell_s^2\int d^2\sigma\l \d_\alpha \d_\beta X^i\d^\alpha \d^\beta X^i\r^2\;.
\]
These terms originate from local geometric invariants, such as $R^2$ and $R_{\alpha\beta}^2$, where $R_{\alpha\beta}$ is the induced curvature of the world sheet metric.
Power counting demonstrates that this term contributes to the spectrum at the order $ \ell_s^6/R^7$, so that all the terms up to $ \ell_s^4/R^5$ are universal. A brute force calculation of all the universal terms is tedious, however, and has not been performed yet.  
Shortly, following \cite{Dubovsky:2012sh}, we will review a shortcut allowing to obtain all the universal $ \ell_s^4/R^5$ terms bypassing a direct calculation.

\begin{figure}[t!]
 \begin{center}
 \includegraphics[width=3.35in]{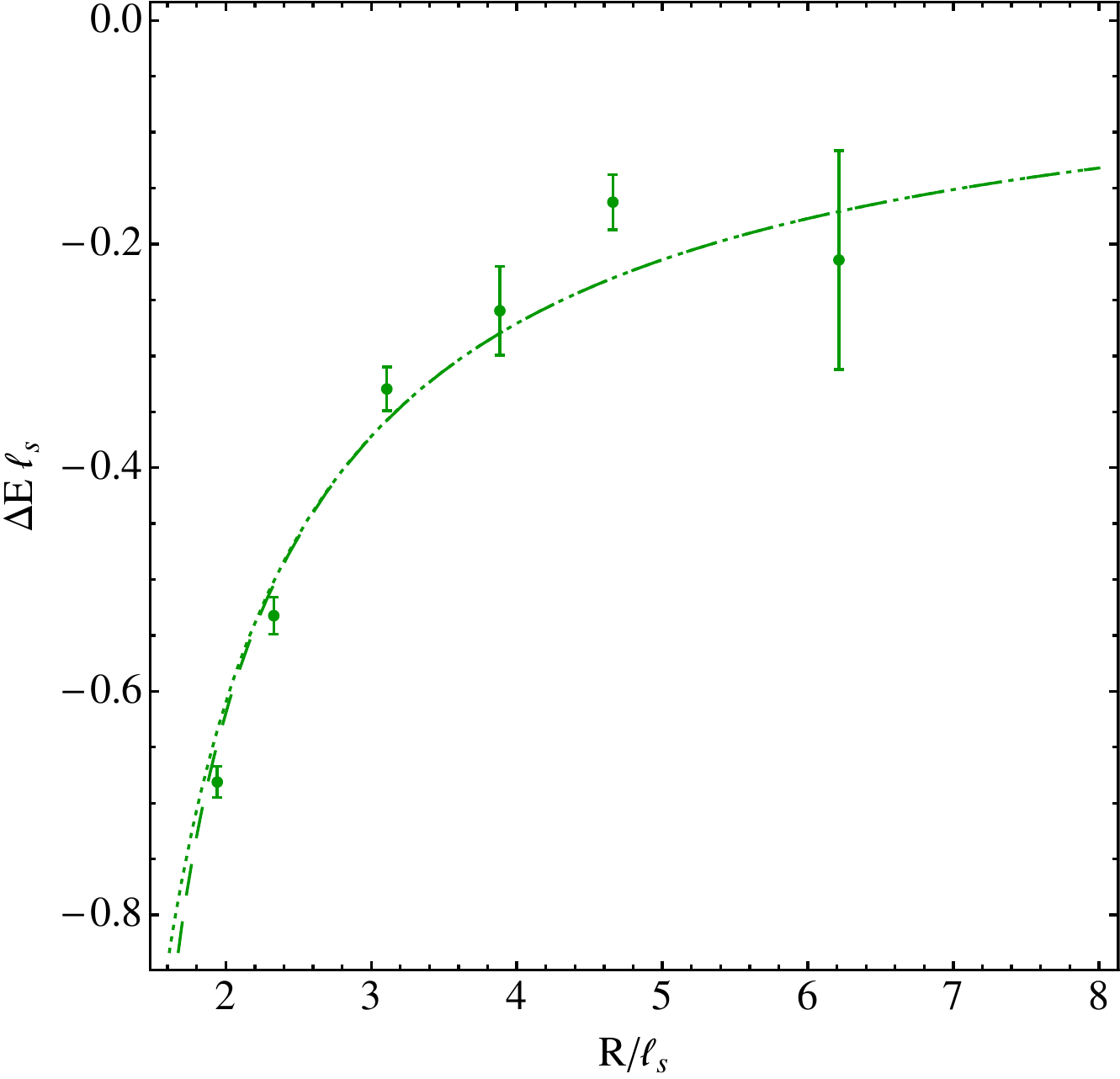}
 \caption{ $\Delta E=E-R/\ell_s^2$ for the ground state of the flux tube. The value of $\ell_s$ was determined from the lattice data. The dotted line shows the prediction of a derivative expansion. The dashed line shows the prediction of the GGRT theory.}
 \label{fig:ground}
 \end{center}
\end{figure}

\begin{figure}[t!]
 \begin{center}
 \hskip 4mm\includegraphics[width=3.2in]{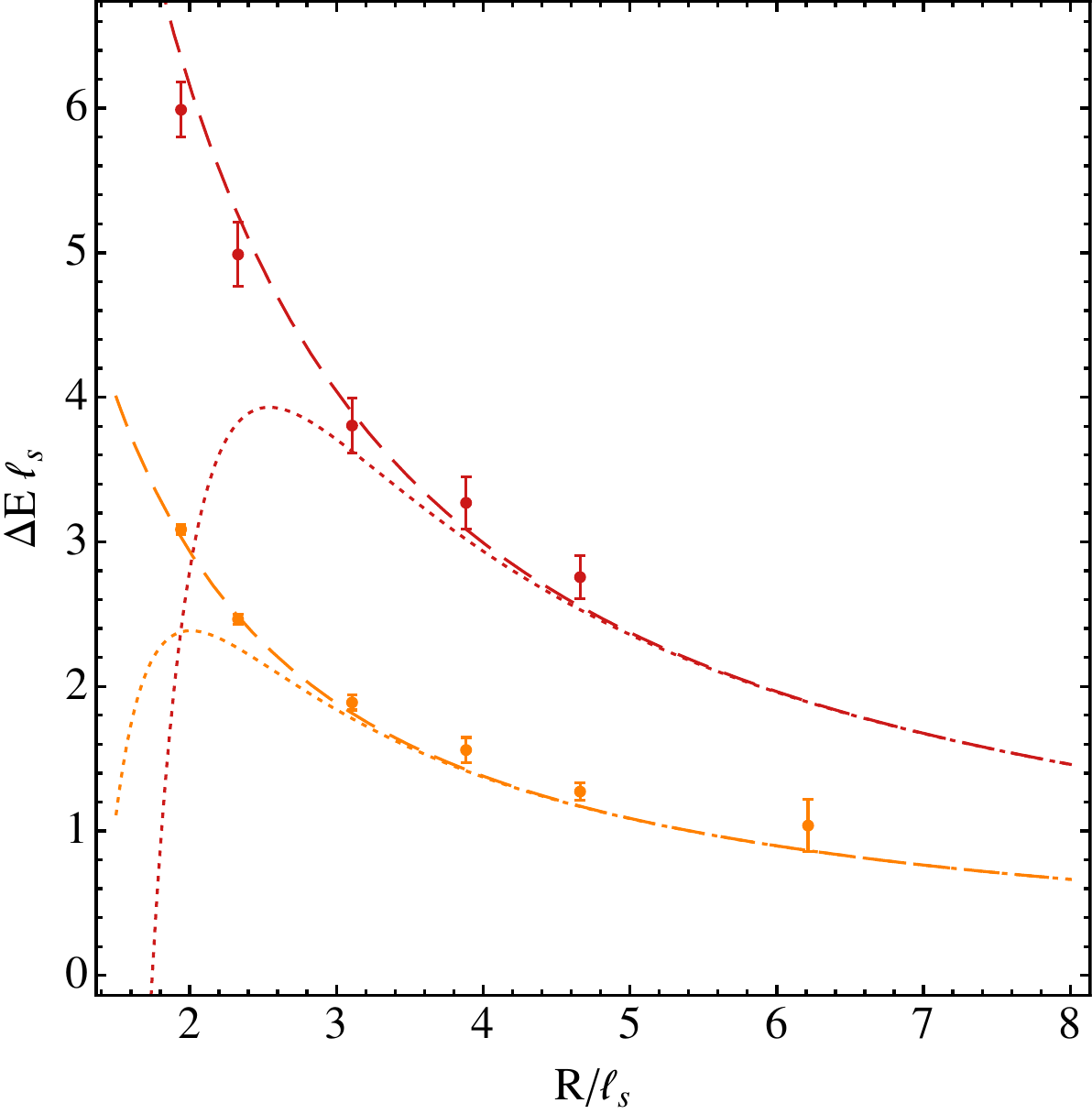}
 \caption{ $\Delta E=E-R/\ell_s^2$ for excited states of the flux tube with one and two units of KK momentum in orange and red, respectively . The dotted lines show the prediction of a derivative expansion. The dashed lines show the prediction of the GGRT theory.}
 \label{fig:oneKK}
 \end{center}
\end{figure}

\begin{figure}[t!]
 \begin{center}
 \includegraphics[width=3.2in]{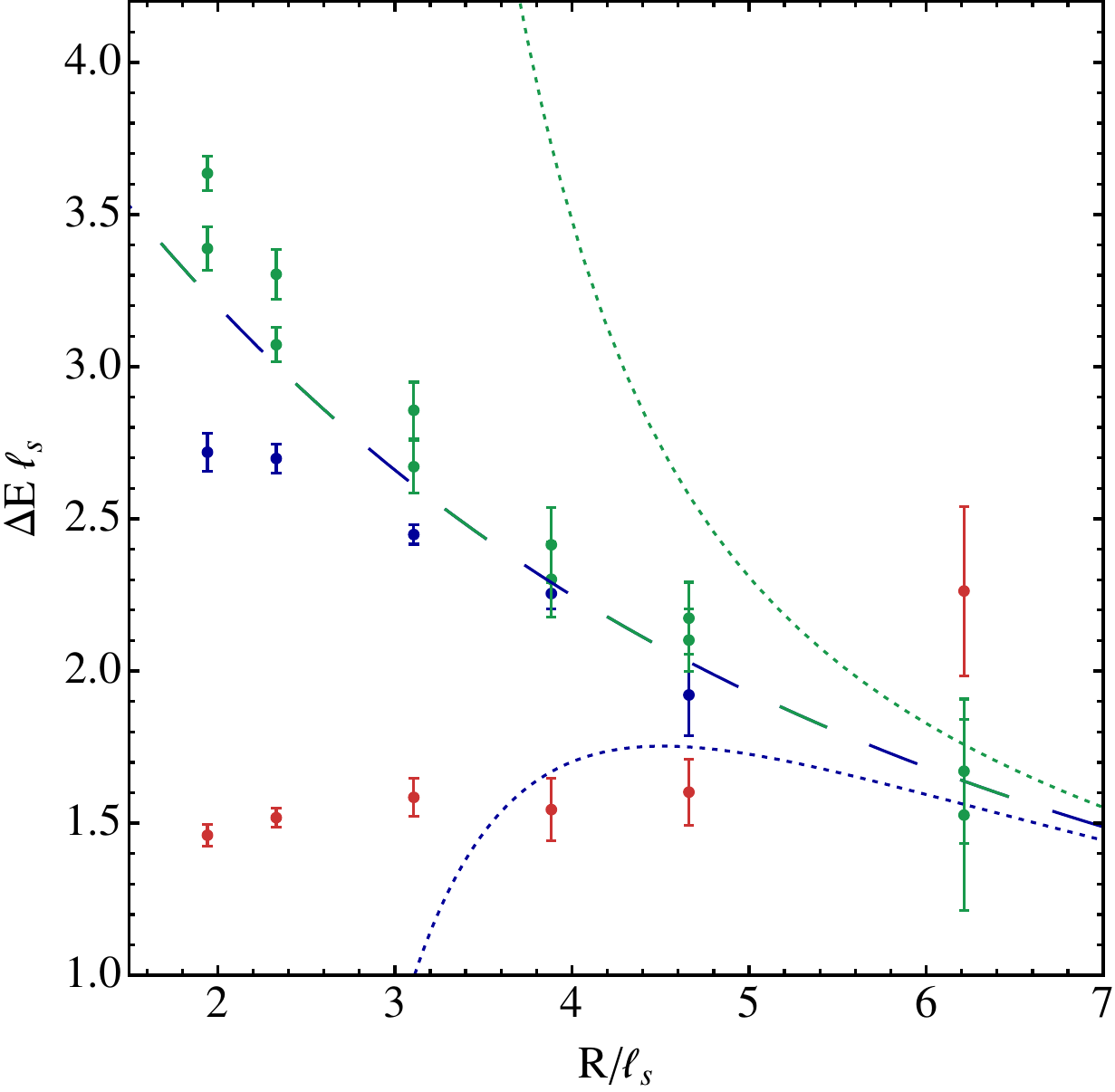}
 \caption{ $\Delta E=E-R/\ell_s^2$ for an excited state with one left- and one right-mover, each with one unit of KK-momentum. The dotted lines show the prediction of a derivative expansion. The dashed lines show the prediction of the GGRT theory. The green color represents a state that is a symmetric tensor with respect to SO(2), the blue color represents the states the scalar with respect to SO(2) and the red data points represent the anti-symmetric tensor with respect to SO(2). All states are predicted to be degenerate in the GGRT theory. In the derivative expansion, the scalar and anti-symmetric tensor are still predicted to be degenerate as indicated by the blue dotted line. The degeneracy with the symmetric state is lifted, which is predicted to have higher energies as shown by the green dotted line.}
 \label{fig:twoKK}
 \end{center}
\end{figure}

Confronting the effective string theory predictions with lattice data for $D=4$ $SU(3)$ gluodynamics leads to several puzzles, as can be seen from Figs.~\ref{fig:ground}, \ref{fig:oneKK} and \ref{fig:twoKK}.
The data points on these plots represent  string energies as a function of the compactification size $R$.  Fig.~\ref{fig:ground} presents the ground state energies, Fig.~\ref{fig:oneKK}
presents states with a single left-moving phonon with different values of the KK momentum, and Fig.~\ref{fig:twoKK} shows a state with one left-moving and one right-moving phonons each with one unit of KK momentum. In the latter case different colors label different two-particle states, classified according to representations of the $O(2)$ group of unbroken rotations in the transverse plane.

In addition, we presented two theoretical expectations for how these energies might look like.
Dotted lines show the sum of universal $ \ell_s^4/R^5$ effective string theory terms. As explained above, these follow from the consistent first principle calculation 
and should agree with the data for sufficiently long strings. 

The second set of theoretical curves, shown as dashed lines, is an {\it ad hoc} spectrum, which is traditionally referred to as the ``free string spectrum" in the lattice community, following \cite{Arvis:1983fp}.
It is obtained by applying the light cone quantization method
of \cite{Goddard:1973qh} to a free bosonic string at $D=4$,
\be
\label{LCspectrum}
E_{LC}(N,\tilde N)=\sqrt{{4\pi^2(N-\tilde{N})^2\over R^2}+{R^2\over \ell_s^4}+{4\pi\over \ell_s^2}\l N+\tilde{N}-{D-2\over 12}\r}\;.
\ee
Here $R$ is the length of the string, $N$ and $\tilde{N}$ are levels of an excited string state, so that $2\pi(N-\tilde{N})/R$ is the total Kaluza--Klein momentum of the state. 
 In what follows we will refer to this spectrum as the GGRT spectrum.
One does not expect this spectrum to match the spectrum of the QCD flux tube. Indeed, as discussed above, non-linearly realized Lorentz
symmetry imposes strong constraints on the properties of the QCD strings. The light cone quantization is famously incompatible  with the target space Poincar{\'e} group away from the critical dimension $D = 26$~\footnote{Another interesting exception is $D=3$, c.f. \cite{Mezincescu:2010yp}.}. Hence, {\it a priori}, one might only expect an agreement with a classical limit of the GGRT spectrum,
in the regime in which the quantum effects can be neglected.

Nevertheless,  as seen from Figs.~\ref{fig:ground}-\ref{fig:twoKK}, the GGRT spectrum surprisingly fits the lattice data better than the perturbative calculations. 
In fact, the situation 
is somewhat different for different classes of states. 
For the ground state,  Fig.~\ref{fig:ground}, both perturbative calculations and 
the GGRT spectrum agree with each other and with the data even for the shortest strings. This is a surprise on its own,
given that the agreement holds even for strings as short as $R=2\ell_s$.

For the purely left-moving states, Fig.~\ref{fig:oneKK}, perturbative calculations agree with the GGRT spectrum and with the data for relatively long strings. For shorter strings perturbative expansion breaks down and the data follow the GGRT prediction.

Finally, for the state with both left- and right-moving phonons,  Fig.~\ref{fig:twoKK}, the perturbative expansion is practically useless in the range of the string length for which lattice data is available. The GGRT approximation provides a reasonable approximation for some of the states (the scalar and  symmetric tensor), while
others (the pseudoscalar) are not explained at all.

These observations taken together provide strong motivation to set up an alternative perturbative expansion with  better convergence properties.
As a first step one would like to understand the physical origin of the GGRT spectrum for $D\neq 26$. As presented at the moment  
it only has the status of an {\it ad hoc} fitting formula.

The answer to this question was given in \cite{Dubovsky:2012sh,Dubovsky:2012wk}. For any value of $D$, the GGRT formula (\ref{LCspectrum}) provides an exact answer for the finite volume spectrum of a certain integrable reflectionless relativistic two-dimensional theory. The exact $S$-matrix of this theory is determined by a two-particle scattering phase shift of the form
\be
\label{GGRTphase}
e^{2i\delta_{GGRT}}=e^{i\ell_s^2 s/4}\;.
\ee
The special role of the  critical dimension $D=26$ (and also of $D=3$, \cite{Mezincescu:2010yp}) is that in this case the theory is both integrable and enjoys a non-linearly realized target space Poincar{\'e} symmetry $ISO(1,D-1)$. Existence of this family of integrable models is not surprising, given that the light cone string quantization provides a direct canonical construction of the corresponding Hilbert space, and does not break the two-dimensional part of the Poincar{\'e} algebra.
However, Lorentz invariance had not been firmly established prior to~\cite{Dubovsky:2012sh,Dubovsky:2012wk}\footnote{We thank Ofer Aharony and Zohar Komargodski for emphasizing this point to us.}. The subtlety is that the conventional light cone quantization is performed in the sector with zero winding number, while the spectrum (\ref{LCspectrum}) arises in the sector with a non-trivial winding. The normal ordering constant in the light cone quantization (which determines the $(D-2)$-term in (\ref{LCspectrum})) is usually fixed by imposing the target space Poincar{\'e} symmetry and it remains unclear what fixes it in the non-critical dimension.

These questions are resolved by applying the TBA method to reconstruct the finite volume spectrum from the $S$-matrix (\ref{GGRTphase}). This exactly reproduces the GGRT spectrum (\ref{LCspectrum}), demonstrating both that the GGRT spectrum
is indeed the finite volume spectrum of a relativistic two-dimensional theory and showing that the normal ordering constant is in fact fixed from the requirement of a two-dimensional Poincar{\'e} symmetry alone in the sector with a non-trivial winding.

This observation turns out to be important for the idea behind the method described in this paper, and to illustrate its power, let us review, following~\cite{Dubovsky:2012sh}, how it allows to derive the universal part of the flux tube spectrum in the conventional $\ell_s/R$ expansion in a simple way. By straightforward perturbative calculation of the scattering amplitudes  one finds that at the level of the Lagrangian the relation between the integrable family of GGRT theories and of the effective theory on the world sheet of an infinitely long relativistic flux tube takes the form 
 \be
 \label{PSrelation}
 {\cal L}_{GGRT}={\cal L}_{NG}+{\cal L}_{PS} +\dots\;.
 \ee
 Here ${\cal L}_{GGRT}$ stands for the Lagrangian of the GGRT theory (determined by the $S$-matrix (\ref{GGRTphase})), ${\cal L}_{NG}$ is the Lagrangian of the relativistic flux tube theory,
 ${\cal L}_{PS}$ is the Polchinski--Strominger (PS) operator \cite{Polchinski:1991ax},
 \be
 \label{PSoperator}
 {\cal L}_{PS}={D-26\over 192\pi}\d_\alpha\d_\beta X^i\d^\alpha\d^\beta X^i\d_\gamma X^j \d^\gamma X^j+\dots\;,
 \ee
  and $\dots$'s stand for higher order terms in the $\ell_s$-expansion. Upon compactification on a circle of circumference $R$, the infinite volume relation (\ref{PSrelation}) implies that up to the order $(\ell_s/R)^3$ the flux tube spectrum coincides with the expansion of the GGRT spectrum. The leading $(\ell_s/R)^5$-difference between the two is given simply by the matrix elements of the PS operator. This is the fastest way to derive the universal perturbative $\ell_s/R$-results presented in 
  Figs.~\ref{fig:ground}-\ref{fig:twoKK}. This general argument agrees with the explicit calculations \cite{Aharony:2011ga} performed for a large set of states in the conformal gauge. 
  
In fact, the flux tube spectrum exhibits an even larger set of universal relations. The relation (\ref{PSrelation})
is a consequence of the universality of the one-loop two-to-two scattering amplitude on the worldsheet of the relativistic flux tube. Power counting demonstrates that actually arbitrary connected one-loop amplitudes are universal and determined solely by the NG part of the action. At finite volume this universality translates into 
relations between energies of different flux tube excitations at higher order in $\ell_s/R$-expansion. These may be checked by inspecting the leading corrections to 
binding energies of different states.

Unfortunately, as discussed above, in spite of this high degree of universality, the conventional $\ell_s/R$-expansion is not very useful for the study of the excited flux tube states  observed in current  lattice simulations, which brings us to the main subject of this paper, the description of an alternative technique based on the thermodynamic Bethe Ansatz.

\section{Finite Volume Spectra From Infinite Volume Scattering}
\label{sec:diagrams}
To find a cure for the bad convergence property of the $\ell_s/R$-expansion, let us first understand the physical origin of the problem.
Why do excited states behave so much worse than the ground state, for which the expansion works extremely well?
The difference between the ground state and the excited states is visible already in the GGRT theory.
As apparent from the expression (\ref{LCspectrum}) for the GGRT spectrum the $\ell_s/R$-expansion for excited states breaks down when $R^2/\ell_s^2 \sim 4\pi(N+\tilde{N})$, which can be a relatively large number. For the ground state, however, the radius of convergence corresponds to $R^2/\ell_s^2=(D-2)\pi/3$, which is much smaller.
Physically, the origin of additional terms of order $2\pi N/R$ in the excited states energies is clear. These are the momenta of free phonons comprising the excited state.
This suggests that it is useful to think of the finite volume energies to be functions of the form
\[
E=\ell_s^{-1}{\cal E}(p_i\ell_s,\ell_s/R)\;,
\]
where $p_i$ are the momenta of individual particles propagating on the world sheet.
The conventional $\ell_s/R$ expansion assumes the free theory answer for $p_i$ and expands the resulting function in $\ell_s/R$.
The key idea of the new method is to calculate the spectrum in such a way that these two functional dependences become disentangled. 

Our previous discussion, most notably the {\it definition} of the GGRT theory by its scattering phase shift (\ref{GGRTphase}) suggests a natural language to achieve this.
One should perform the calculation of the finite volume spectrum in two distinct steps.
One first calculates the (infinite-volume) $S$-matrix and then proceeds towards extracting the finite-volume spectrum from this $S$-matrix. The first step corresponds to the perturbative expansion in $p_i \ell_s$ and because of the usual analytic properties of the $S$-matrix turns out to be convergent even for momenta which are not particularly small. 
%however, because the spectrum doesn't explicitly depend on $p_i$'s this property gets  

Even though it is widely believed that the $S$-matrix of a quantum field theory
uniquely determines its finite-volume properties, the prescription for the second step is not known in general. However, it is understood in two circumstances. 

For massive theories below the particle production threshold there is a perturbative procedure first implemented by L\"uscher \cite{Luscher:1986pf} commonly used in lattice calculations. There is no principal obstruction to extend this technique above the inelastic threshold and multichannel generalizations of L\"uscher formulas are being developed (see, e.g., \cite{Hansen:2012tf,Hansen:2013dla}). One of the major challenges (at least at the technical level) within this approach is to calculate winding corrections, coming from virtual particles traveling around the compact dimension. In massive theories these are exponentially suppressed, and usually are either neglected, or calculated by accounting for the lowest order winding contributions.
In a massless theory, like the effective string theory, more care is needed because the winding corrections are only power law suppressed.
 
For two-dimensional integrable theories there is an exact (non-perturbative) method for calculating the finite volume
spectrum known as the Thermodynamic Bethe Ansatz (TBA) \cite{Zamolodchikov:1989cf,Dorey:1996re}. Even writing down the complete set of equations, especially for excited states, is in general quite non-trivial and usually involves some amount of guesswork.
However, there is a special class of reflectionless integrable scattering, where the excited states TBA appears to take a simple universal form \cite{Teschner:2007ng,Dubovsky:2012wk}. The GGRT model belongs to this class and the corresponding set of excited states TBA equations is known exactly.

The world sheet theory of flux tubes does not have a mass gap and is not integrable.
However, its leading order scattering amplitudes (in $p\ell_s$ expansion) coincide with those of the GGRT theory.
At the next-to-leading order relativistic effective strings deviate from the GGRT for general D, and reflections and annihilations appear at this order. However, in D=4 they still take a special form for which it is possible to write down the full set of excited state TBA equations. This will be our starting point for the analysis of the flux tube spectra observed on the lattice.
As we will see, this method provides much better control of the spectra than the conventional $\ell_s/R$-expansion, and makes it clear that the minimal effective string theory needs to be extended to explain the lattice data. The extension will be incorporated in the TBA equations perturbatively.

\subsection{Thermodynamic Bethe Ansatz for reflectionless scattering}
\label{sec:TBA}

Let us review the basics of the thermodynamic Bethe Ansatz. For now we consider massless theories with integrable reflectionless $S$-matrices with any number of particle species. By integrability we mean that in every scattering process the number of particles is conserved, the final particles have the same momenta as the initial ones and the absence of reflections implies that the final distribution of flavors coincides with the initial one. 
%TBA is also applicable to massive integrable theories which allow particles to exchange flavor indices in course of scattering, however massless  relativistic integrable theories cannot have reflections as follows from a certain consistency condition on the $S$-matrix - the Yang-Baxter equation. In particular this means that the flux tube theory in fact cannot be integrable because reflections appear in it at the one-loop level. 
Integrability implies  that the $S$-matrix element for scattering of $n$ left- and $m$ right-moving particles is equal to the product of $n \times m$ pairwise $S$-matrices. Every $2\to2$ $S$-matrix element in every flavor channel must be just a number with absolute value 1, as demanded by unitarity,
$$
S_{ab}^{cd}=\delta_a^c \delta_b^d \e^{2 i \delta_{ab}}\;.
$$
%Since the particles are massless it is convenient to divide them
%into left and right movers and to label them by the spacial component of the momenta which we choose to %be positive for both: $p^{(j)}_{il}$, $p^{(j)}_{ir}$. 
%For the sake of generality for now we also keep track of the flavor index in the brackets. Then the momenta should satisfy the periodicity condition modified by the interactions:

The TBA allows to extract the finite volume spectrum of the theory  from the phase shifts $\delta_{ab}$. There are three key ideas underlying this method. The first is called Asymptotic Bethe Ansatz (ABA). It is a set of algebraic equations which gives the spectrum in the approximation where the contributions from virtual particles traveling around the ``world'' are neglected. The ABA equations are discussed in more details in the subsection \ref{sec:ABA} together with their derivation.

The second idea is the following: instead of considering the theory in the finite volume $R$ and at zero temperature, one considers the theory in which time and space directions are interchanged. Consequently this theory appears to live in an infinite volume but at a finite temperature $T=1/R$. For a relativistic theory the space-time interchanged (``mirror") theory coincides with the initial one. The asymptotic Bethe Ansatz becomes exact in the thermodynamic limit
 and  takes the form of certain integral equations which allow to find the free energy density $f(T)$ in the mirror interchanged theory. The functional integral representation of the partition function implies that it is related to the ground state energy of the initial theory as 
 \[
 E_0(R)=R f(1/R)\;.
 \]
%functional integral-?
To calculate the energy of  excited states the third idea is needed. The prescription is to deform the contour in the integral equations used for the calculation of the ground state energy in a certain way \cite{Dorey:1996re}. Although the derivation of this procedure for the general case is not yet known, there is a rigorous mathematical proof of the resulting TBA equations for certain integrable theories, such as the sinh-Gordon model  \cite{Teschner:2007ng}.
For the GGRT theory (the case we are mainly interested in here) enough rather non-trivial checks were performed \cite{Dubovsky:2012wk} to be certain that the method can be safely applied. In addition to this, in section \ref{sec:diagTBA}, we will provide partial diagrammatic intuition behind the TBA equations. 

We now turn to presenting the TBA equations themselves. There are two contributions to the energy of a state in this formalism. First, there are ``real" particles, with (positive) momenta equal to  $p_{li}$ and $p_{r i}$ for left and right movers present in the state.
In addition, there is a  ``thermal bath" of particles with pseudo-energies $\eps^a_l(q)$ and $\eps^a_r(q)$  for left- and right-moving components of the bath\footnote{As before, we denote the energy as $\Delta E$, as a reminder that the full energy $E$ contains in addition the classical string tension contribution $R/\ell_s^2$.},
\begin{gather}
\Delta E=\sum\limits_i p_{li}+\sum\limits_i p_{ri}+\frac{1}{2\pi }\sum\limits_a \int_0^\infty dq\ln\left(1-e^{-R\eps^a_l(q)}\right)+\frac{1}{2\pi }\sum\limits_a\int_0^\infty dq\ln\left(1-e^{-R\eps^a_r(q)}\right)\,.
\label{TBAE}
\end{gather}
The thermal bath contribution is responsible for winding corrections and indeed has a thermal origin from the viewpoint of the mirror theory.
To distinguish  thermal particles from the real ones we will be denoting the momenta of the former by $q$. The index $a$ labels a flavor.
The momenta $p_i$ label the state. The case without real particles naturally corresponds to the vacuum state. 

The real particle momenta $p$'s and the pseudo-energies $\epsilon(q)$'s are determined from solving the TBA set of integral equations. These consist of two groups of equations. First, there are generalized quantization conditions for the real momenta
\be
\label{TBApl}
p_{li} R+\sum\limits_j 2\delta_{a_i a_j}(p_{li},p_{rj})-i\sum\limits_b\int_0^\infty \frac{dq}{2\pi}\frac{d\,2\delta_{a_ib}(ip_{li},q)}{dq}\ln\left(1-e^{-R\eps^b_r(q)}\right)=2 \pi N_{i}
\ee
\be
\label{TBApr}
p_{ri} R+\sum\limits_j 2\delta_{a_j a_i}(p_{ri},p_{lj})+i\sum\limits_b\int_0^\infty \frac{dq}{2\pi}\frac{d\,2\delta_{b a_i}(-ip_{ri},q)}{dq}\ln\left(1-e^{-R\eps^b_l(q)}\right)=2 \pi \tilde {N_{i}}
\ee
In the absence of interactions, $\delta=0$, these reduce to the free theory quantization conditions for a set of particles on a circle.
For an interacting theory, the quantization condition is modified for two reasons. First, pairwise interactions between real particles explain the appearance of the corresponding phase shifts in (\ref{TBApl}), (\ref{TBApr}) (we will explain the origin of this effect in section~\ref{sec:ABA} in detail).
Second, there are integral contributions, which account for winding corrections. Imaginary momenta appearing in (\ref{TBApl})-(\ref{TBAepsr}) came from performing the double Wick rotation to the mirror theory.
However, the crossing symmetry which in terms of the phase shift reads $\delta(p_l,-p_r)=\delta(-p_l,p_r)=-\delta(p_l,p_r)$ guarantees that the equations are actually real.
We did not make use of it to simplify the equations and get rid of $i$'s, because the crossing symmetry will get modified in the presence of annihilations, which will be discussed later.

Finally,
%For our purposes it is enough to consider the situation where 
%diagonal-??
the pseudo-energies satisfy the following TBA constraints
\be
\label{TBAepsl}
\eps^a_l(q)=q+\frac{i}{R}\sum\limits_i 2\delta_{ab_i}(q,-i{p}_{r i})+\frac{1}{2\pi R}\sum\limits_b\int_0^\infty dq'\frac{d\,2\delta_{ab}(q,q')}{dq'}\ln\left(1-e^{-R\eps_r^b(q')}\right)
\ee
\be
\label{TBAepsr}
\eps_r^a(q)=q-\frac{i}{R}\sum\limits_i 2\delta_{b_i a}(q,i{p}_{li})+\frac{1}{2\pi R}\sum\limits_b\int_0^\infty dq'\frac{d\,2\delta_{ba}(q,q')}{dq'}\ln\left(1-e^{-R\eps^b_l(q')}\right)
\ee
For the  GGRT phase shift $2\delta_{a_ib_j}=\ell_s^2 p_{li} p_{rj}$ it is straightforward to solve the full TBA system (\ref{TBApl})-(\ref{TBAepsr}) 
analytically resulting in (\ref{LCspectrum}).
Note that in the massive sinh-Gordon model the full TBA system takes the same form \cite{Teschner:2007ng}, strongly suggesting that this form should be universal for reflectionless scattering.
The full set of TBA equations has a rather intimidating appearance,
 however as we just explained, the major complications come from winding corrections.
Dropping them results in the Asymptotic Bethe Ansatz (ABA) equations, which are known as the L\"uscher formula in the context of lattice calculations,
\be
\label{ABA}
p_{l(r)i} R+\sum\limits_j 2\delta_{a_i a_j}(p_{l(r) i},p_{r(l) j})=2 \pi N_{i}\;.
\ee
For the massive case all the integral terms are suppressed as $\exp(-\mu R)$, where $\mu$ is the mass gap, as is natural to expect for winding corrections. 
In our case the winding corrections are only power-law suppressed, and we have to pay more attention to them. However, as we will see,
 for the values of $R$ we consider, the main effect is still coming from the asymptotic part of the Bethe Ansatz. We will explain the reason for 
 this  in section~\ref{sec:IRwind}.
 
\subsection{Asymptotic Bethe Ansatz}
\label{sec:ABA}
In this section we sketch a simple derivation of the multi-channel generalization of the asymptotic Bethe Ansatz equatios. It is certainly not new.
One of the reasons to present the ABA derivation here is to stress that the 
logic underlying this derivation does not directly rely on integrability. 
In particular, we allow for non-diagonal scattering, so that the amplitude is no longer reflectionless.
 Conceptually,  there appears to be no obstacle in generalizing the  ABA to accommodate  particle production. For example to account for the $2\leftrightarrow4$ processes one should add  matrix elements mixing two- and four- particle states. In the case at hand, however, these processes are suppressed at low energies. 
So in what follows we neglect these effects and assume that the $2\to2$ part of the $S$-matrix $S_{ab}^{cd}$ is unitary.

 First let us consider two particles in an infinite volume, the first one moving to the right and the second one moving to the left. The basis for the in-states is formed by  $\ket{p_r,a\,;p_l,b }$ and the wave function of a generic state is 
  defined as
\be
\psi^{ab}(x_1,x_2)=\bra{0} \phi^a(x_1) \phi^b(x_2) F^{cd}\ket {p_r,c\,; p_l,d}\;,
\ee
where the field operators are taken at equal time, $F^{cd}$ denotes the flavor wave function, and we suppressed the time dependence.
Strictly speaking, our discussion assumes  that the states are taken to be wave packets, but to keep the formulas short
we do not write this explicitly.
When particles are far apart they do not interact with each other, the energy of the state is given by $|p_l|+|p_r|$ and the wave function is just a product of two plane waves. Thus in the region $x_1 \ll x_2$ the wave function consists of two contributions: either the first particle is found at $x_1$ and the second at $x_2$ --- before they scattered, or the second particle is found at $x_1$ and the first at $x_2$. In the latter case the particles have to scatter before they reach their positions.
As a result the total 
wave function in this region takes the form
\be
\psi^{ab}(x_1\ll x_2)=\e^{i p_r x_1+i p_l x_2} F^{ab}+\e^{i p_l x_1+i p_r x_2} S_{cd}^{ba}F^{cd}.
\ee
The same reasoning applied in the region  $x_1\gg x_2$ gives
\be
\psi^{ab}(x_1\gg x_2)=\e^{i p_l x_1+i p_r x_2} F^{ba}+\e^{i p_r x_1+i p_l x_2} S_{cd}^{ab}F^{cd}.
\ee
Now we will consider this state in a finite volume and impose the corresponding periodicity condition. To achieve  this,  let us consider  $x_1$ and $x_2$ such that $x_1 \ll x_2 \ll x_1+R$. Then the periodicity of the wave function $\psi(x_1,x_2)=\psi(x_1+R,x_2)$ demands 
\be
\label{period}
\e^{ip_rR}S_{cd}^{ab}F^{cd}=F^{ab}
\ee
All other periodicity conditions are equivalent because the total momentum $p_1+p_2$ is quantized in units of $2\pi/R$.
Eq (\ref{period}) has solutions iff 
\be
\label{Luscher}
\det\l e^{ip_rR}S_{cd}^{ab}-\delta_c^a \delta _d^b \r =0,
\ee
where $(ab)$ as well as $(cd)$ should be treated as a single matrix index when the determinant is taken.
This is the multi-channel generalization of the L\"uscher formula, which imposes the relation between the $S$-matrix and the allowed momenta of particles in the finite volume. If the $S$-matrix is known it allows to find the energy spectra, given by $|p_l|+|p_r|$. Conversely, if the spectra as functions of $R$ are known one can reconstruct the $S$-matrix. 

It is straightforward to extend this derivation for mulitparticle states in integrable theories. In particular, for reflectionless scattering one immediately arrives at (\ref{ABA}).
As we already said, there appears no fundamental obstacle to extend these arguments for non-integrable theories, even though obtaining the explicit equations is likely to be quite challenging
due to inevitable mixing between states with different number of particles.

Anticipating what will happen later, let us point out one of the main reasons why the TBA technique  displays better convergence than the conventional $(\ell_s/R)$-expansion. As follows from the ABA equations (\ref{ABA}) the actual momenta $p_i$ of interacting phonons are smaller than the free theory value $2\pi N_i/R$, if the phase shift $\delta$ is a growing function of the momentum. Given that the perturbative parameter for the low energy expansion is  $p_i\ell_s$,
accounting for this effect improves the convergence properties of the expansion.

%Let us mention that conceptually there is no obstacle in generalizing the asymptotic Bethe Ansatz to the case where the particle production is present. For example to account for the $2\leftrightarrow4$ processes one should add the four-particle states to the basis and add the matrix elements mixing them with the two-particle states. In the case at hand these processes are suppressed at low energies, so the derivation above will be sufficient for our purposes, however, unlike the part related to winding corrections, ABA does not require integrability. This is why we did not meet any difficulties in adding the reflections to the scattering of massless particles, which is inconsistent with integrability.

\subsection{Towards a diagrammatic interpretation of TBA}
\label{sec:diagTBA}
From the presented derivation it is clear that the winding corrections are absent in the ABA system 
(\ref{Luscher}) because we did not take into account virtual quanta propagating around the world. For the GGRT theory
these are accounted for by the  ``thermal" contributions  in (\ref{TBAE}) and (\ref{TBApl}), (\ref{TBApr}) together with a set of integral equations 
(\ref{TBAepsl}), (\ref{TBAepsr})
for pseudo-energies. 

These equations were obtained following the idea pioneered in \cite{Dorey:1996re}.
The starting point are the ground state TBA equations derived in \cite{Zamolodchikov:1989cf}. These are (\ref{TBAE}) and (\ref{TBAepsl}), (\ref{TBAepsr})
without any real particle contributions.
The idea of  \cite{Dorey:1996re} is that the ground state energy as a function of a sufficiently large set of external parameters allows the reconstruction of the full set of excited states energies by analytic continuation in the parameters and exploiting the monodromies the equations and solutions undergo when circling singularities in the complex plane. 

To arrive at the excited states TBA equations for the GGRT model, one may for example introduce chemical potentials $\mu_{l(r)}^a$ for the number of phonons. These are incorporated  by shifting the pseudo-energies
\[
\epsilon^a_{l(r)}\to \epsilon^a_{l(r)}+\mu_{l(r)}^a
\]
in the thermal integrals in 
(\ref{TBAepsl}), (\ref{TBAepsr}). As a result of an analytical continuation along a contour in the complex plane of $\mu$'s, which starts and ends 
at $\mu=0$, the integrals may pick up extra contributions from circling around the branch points of the logarithm. These give rise to the contributions in (\ref{TBAepsl}), (\ref{TBAepsr}) corresponding to real particles. The generalized ABA equations (\ref{TBApl}), (\ref{TBApr}) determine  the positions of the singularities.
Particle  number simply counts  how many times different singularities were circled.

Unfortunately, there is still an ambiguity left in this prescription concerning the correct direction for circling around the singularities (the one corresponding to positive particle numbers $N_i$). This may be fixed by requiring that one reproduces the correct result in the free theory limit, $\ell_s\to 0$. 

This line of reasoning leads to the correct result for the excited states TBA. Nevertheless, it is tempting to look for a diagrammatic understanding of how the excited states TBA arises. In particular, one may hope to see that it corresponds to a certain resummation of the conventional perturbative expansion, which would help to illuminate the origin of the better convergence of the TBA method.
Some insight into this issue was given in \cite{Luscher:1985dn,Luscher:1986pf}  (see \cite{Janik:2010kd} for a review and generalization to an arbitrary dispersion relation). However, the proposed diagrammatic method corresponds to an expansion in winding corrections or $\exp(-mR)$ because massive particles were considered. 
Since winding corrections in massless theories are only power law suppressed, this expansion does not provide a good approximation. This motivates us to look for an alternative resummation of Feynman diagrams.

At this point we do not have a complete solution to this problem, but instead merely report on partial progress in this direction.
First, recall that even though our theory is massless and winding corrections are not suppressed exponentially, numerically they are nevertheless small for values of $(D-2)$ of interest.
We already mentioned the reason for this at the end of section \ref{sec:TBA} and will illustrate this point numerically later on.
This suggest an iterative solution of the TBA equations, in which one first ignores the integral parts, finds corresponding $\eps$'s and $p$'s and then solves the integral equations iteratively.

Note that this expansion is different from the expansion in winding number mentioned above. The latter corresponds, roughly, to expanding the thermal TBA logarithms in a series of exponential terms $\e^{-nR\epsilon}$.

 To see that the convergence of this method is good at least for the GGRT theory, notice that that to leading order it corresponds to the expansion of the square-root formula (\ref{LCspectrum}) in a formal parameter $(D-2)$, and the latter is convergent for any state for the values of $R$ and $(D-2)$ we consider.
For the ground state, the $(D-2)$ expansion is equivalent to the $\ell_s/R$ expansion, but they behave differently for all excited states. For instance, completely neglecting the $(D-2)$-contributions results in the following ABA spectrum for the GGRT theory,
\be
E_{cl}(N,\tilde{N})=\ell_s^{-1}\sqrt{{R^2\over\ell_s^2}+{4\pi^2\ell_s^2(N-\tilde{N})^2\over R^2}+{4\pi }\l N+\tilde{N}\r}\;.
\label{1appr}
\ee
This coincides with the spectrum of the classical string. 
%thus it's diagrammatic interpretation is just the usual perturbative one. The first term corresponds to sum of one-loop diagrams. Note, however, that beyond the first step it is a coincidence valid for the GGRT theory only, since the expansion of the phase shift for this theory is trivial.
In the rest of this section we demonstrate how the first term of the $(D-2)$ expansion of the spectrum (\ref{LCspectrum}) arises in the diagrammatic language.

We will organize the calculation in the following way. One starts with a set of particles, corresponding to a chosen state, with momenta determined by the
ABA quantization conditions. At this stage winding corrections are not included yet, so 
it is appropriate to think of this state as a ``gas" in an infinite volume albeit in a very special state in which all particles have the same momentum.
The leading winding corrections then take the form of conventional bubble diagrams with the propagator taken to be the one for fluctuations around this gas.

\begin{figure}[t!]
 \begin{center}
 \includegraphics[width=6.0in]{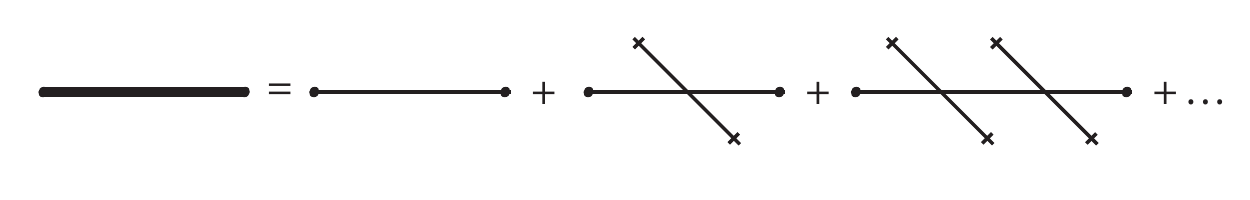}
 \caption{Propagator for a virtual quantum in the presence of the real left-moving particles indicated in the figure by crosses.}
 \label{fig:gas}
 \end{center}
\end{figure}

\begin{figure}[t!]
 \begin{center}
 \hskip 4mm\includegraphics[width=3.2in]{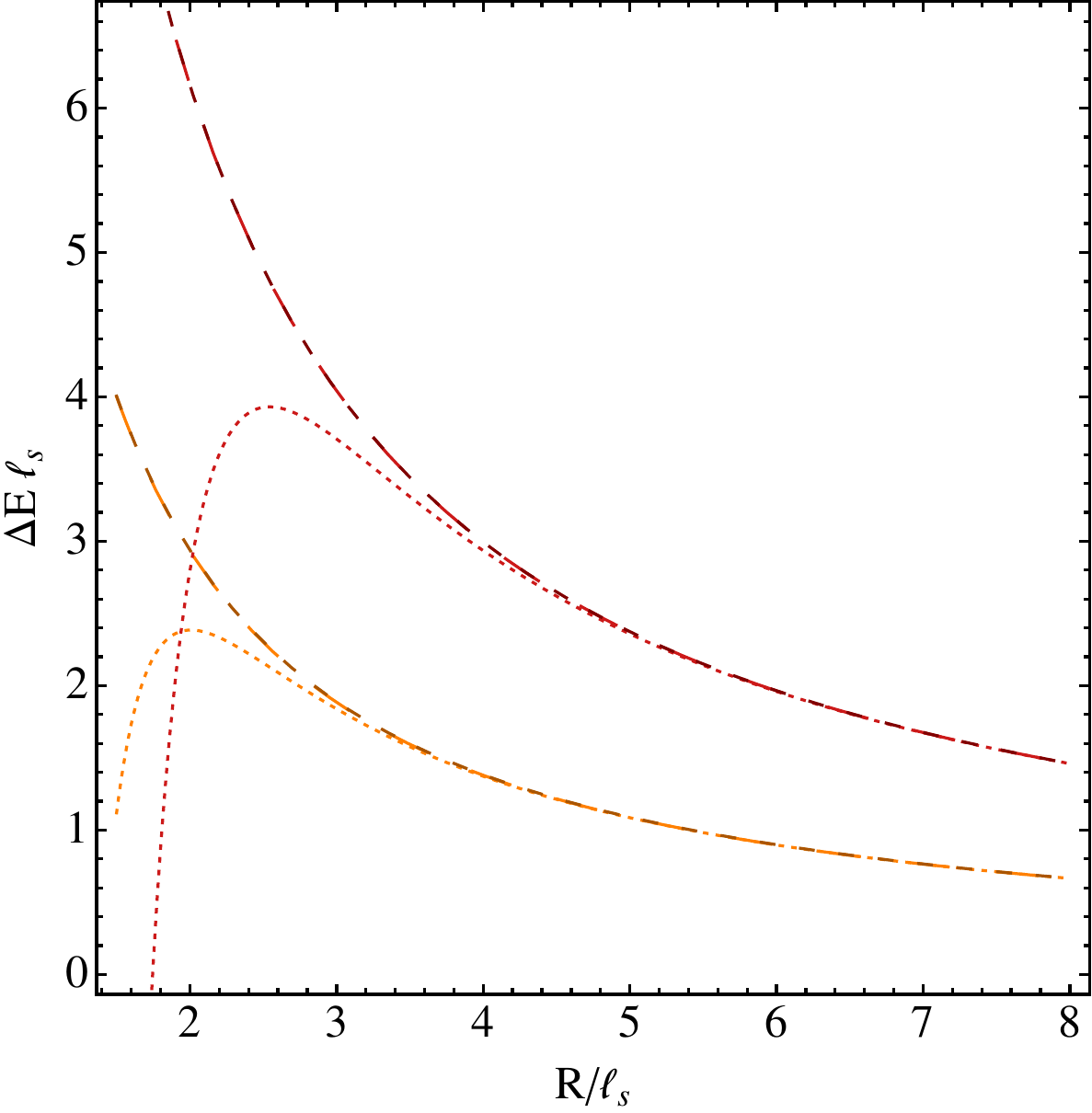}
 \caption{ $\Delta E=E-R/\ell_s^2$ for excited states of the GGRT theory with one and two units of KK momentum in orange and red, respectively. The dotted lines show the prediction of a derivative expansion, the longer dashes show the prediction of the GGRT theory, and the shorter darker dashes represent our diagrammatic approximation. The diagrammatic approximation and the exact result are virtually indistinguishable.}
 \label{fig:gasTBAl}
 \end{center}
\end{figure}

To illustrate how this works in practice, let us first consider a state on a circle with a single left-moving phonon.
The ABA quantization is equivalent to the free one in this case, so the momentum of the particles in the gas is 
 $p_l=2\pi N/R$.
 It is convenient to introduce also the parameter $\alpha_l=\ell_s^2 p_l /R$; physically this is the energy density of the gas (in string units).
 Consider a probe particle with a momentum $q_\alpha$, propagating through the gas. 
To calculate the dressed propagator for this particle one needs to resum the diagrams represented on Fig.~\ref{fig:gas}.
In terms of the momentum expansion, we restrict ourselves to the leading term. So only tree level diagrams are taken into account.  All one-particle irreducible diagrams containing more than two gas insertions vanish in this case, so we obtain the propagator exactly to all orders in $\alpha_l$ as a geometric series,
\begin{gather}
G(q)=\frac{i}{(q_0^2-q_1^2)}+\frac{i}{(q_0^2-q_1^2)}{{i\cal M}\over 2p_l R} \frac{i}{(q_0^2-q_1^2)}+\ldots\;,
\end{gather}
where ${\cal M}$ is the forward scattering amplitude for the scattering of the virtual particle off a phonon in the gas,
\[
{\cal M}= 2\ell_s^2 p_l^2 (q_0+q_1)^2\;,
\]
and the factor of $1/R$ stands for the number density of phonons.
By calculating the geometric series we obtain
\be
\label{Gq}
G(q)=
\frac{i}{(q_0+q_1)\l q_0-q_1+(q_0+q_1) \alpha_l \r}\;.
\ee
Since left-movers do not interact with each other, the dispersion relation for a left-moving quantum does not get modified in a purely left-moving gas. On the other hand, the right moving probe is slowed down by interactions and its dressed dispersion relation is 
\[
q_0={1-\alpha_l\over 1+\alpha_l}q_1\;.
\]
Note that for $\alpha_l>1$ a ``right-mover" gets carried away by the gas and actually propagates to the left.
It is now straightforward to construct the quadratic effective action reproducing the propagator (\ref{Gq}),
\be
\label{quadaction}
S_{eff}=  \int d^2 \sigma \,\l \frac{1+\alpha_l}{2} \dot{ x }^i
  \dot{x}^i+
  \frac{\alpha_l}{2} \dot{x}^i x^{i\prime}-\frac{1-\alpha_l}{2}x^{i\prime} x^{i\prime}\r\;.
\ee
Now, following the logic outlined above,
we calculate the energy of the state on a circle as the sum of the energy of the real left-moving particle, $p_l$ 
and the winding contribution. At leading order the latter is the ground state energy of the free theory with the action (\ref{quadaction}).
It can be calculated either using ground state TBA (for the mirror theory), or directly.
Proceeding in the direct way, we calculate the energy as the expectation value of the Hamiltonian corresponding to (\ref{quadaction}),
\begin{gather}
\langle H_{eff}\rangle=  \left\langle \int_0^R d\sigma \, \frac{1+\alpha_l}{2} \dot{ x }^i
  \dot{x}^i + \frac{1-\alpha_l}{2}x^{i\prime} x^{i\prime}\right\rangle
  \end{gather}
By making use of the Poisson summation formula we write the result as a sum over windings,
\be
\langle H_{eff}\rangle
=\sum_{n\neq 0} \int \frac{d^2 q}{(2 \pi)^2} \frac{i R \l (1+\alpha_l)q_0^2+(1-\alpha_l) q_1^2 \r \e^{i q_1 n R}}{2(q_0+q_1)( q_0-q_1+(q_0+q_1)\alpha_l)}\;,
\ee
where we dropped the zero winding contribution.
After performing the Wick rotation for $q_0$ one closes the $q_1$-contour and takes the $q_1$-integral
by residues. The resulting total energy of the state is 
\begin{gather}  
\Delta E=p_l+\frac{D-2}{2\pi }\l
\int_0^\infty dq\ln\left(1-e^{-Rq}\right)+\int_0^\infty dq\l1-\alpha_l\r \ln\left(1-e^{-Rq(1+\alpha_l)}\right)\r= \nonumber
  \\ =2\pi N/R-(D-2)\frac{\pi R}{6(R^2+2\pi N \ell_s^2)}
  \label{Dm2L}
 \end{gather}
This is exactly what one gets from the first iteration of solving integral equations (\ref{TBAE})--(\ref{TBAepsl}) perturbatively in winding contributions, 
or equivalently by expanding the exact answer (\ref{LCspectrum}) in $(D-2)$. 
Unlike the $\ell_s/R$-expansion,  the approximation (\ref{Dm2L}) provides an extremely accurate estimate for the exact result down to $R\sim\ell_s$, see Fig.~\ref{fig:gasTBAl}.

Furthermore, just like the exact result, the expression (\ref{Dm2L}) turns into a series with a rather small radius of convergence, when expanded in $\ell_s/R$.
The above derivation sheds light on the physical origin of this behavior and on the nature of the improvement achieved by the TBA method.
The singularity in  (\ref{Dm2L}), which determines the radius of convergence of the $\ell_s/R$-expansion corresponds to $\alpha_l=1$, when
both modes become left-movers. Using the dressed propagator in our calculation allows to avoid  spurious singularities associated with this effect.

\begin{figure}[t!]
 \begin{center}
 \hskip 4mm\includegraphics[width=3.2in]{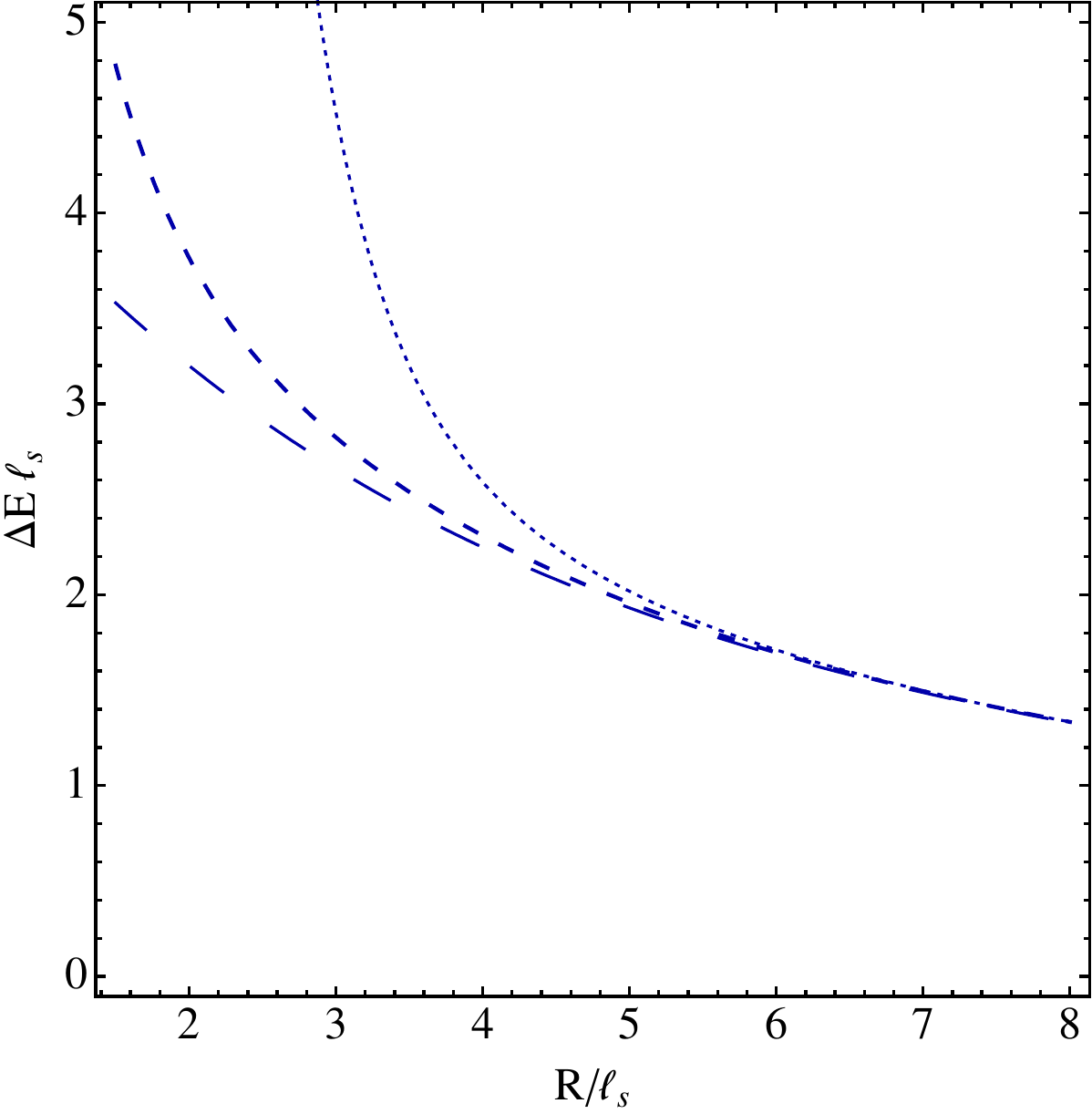}
 \caption{ $\Delta E=E-R/\ell_s^2$ for excited states of the GGRT theory with one a left- and right-moving phonon each with one unit of KK momentum. The dotted lines show the prediction of a derivative expansion, the longer dashes show the prediction of the GGRT theory, and the shorter darker dashes represent our diagrammatic approximation. }
 \label{fig:E11diag}
 \end{center}
\end{figure}

The same reasoning can be applied to all states which contain only left-moving excitations.
In principle, there is no obstacle to extend the same logic to the states containing both left- and right-moving phonons.
One modification in this case is that the momenta of the particles in the gas $p_l$ and $p_r$ are not given by the free quantization condition anymore, rather they are solutions of the L\"uscher equation (\ref{ABA}). 
The difficulty now, however, is how to obtain the result at all orders in the particle's energy densities $\alpha_{l(r)}=\ell_s^2 p_{l(r)} /R$. 
The reason is that now left- and right-moving particles interact with each other, so there are non-vanishing one-particle irreducible diagrams with more than four outgoing legs contributing to the dressed propagator.
It may be possible to sum all these diagrams for the NG action at least at the tree level, but we leave this for future work.
Instead, let us present the perturbative result in  $\alpha$'s, accounting only for the four-particle interactions, as before.
 This leads to the following dressed propagator for a probe particle,
\be
G(q)=\frac{-i}{q_0^2-q_1^2+(q_0^2+q_1^2)\alpha_l+(q_0^2-q_1^2)\alpha_r{R}}\;.
\ee
As expected, the dispersion relation for both left- and right-movers is modified in this case.
Calculation similar to the one we did for  the purely left-moving state results in the following expression for the energy 
at the leading order in $\alpha$'s,
\be
\Delta E=p_l+p_r- (D-2)\frac{\pi\l1-\alpha_l-\alpha_r\r}{6 R}
\ee
As illustrated in Fig.~\ref{fig:E11diag},  this leads to a significant improvement when compared to the naive $\ell_s/R$ expansion, but still is not accurate at small
radii, where the energy densities $\alpha_{l(r)}$ become large and multiparticle interactions must be included.

We feel the above perturbative examples serve well the purpose of illustrating the physics underlying the TBA method. It is an interesting open question whether they can be pushed to higher order. We already mentioned that already at the tree level one needs to learn how to resum an infinite number of tree-level diagrams. 
But one may also try to be more ambitious and push the matching calculation resulting in the effective action (\ref{quadaction}) to higher order. It would be interesting to study whether this method allows to reproduce the full TBA system in the $(D-2)$-expansion, or whether new physical ingredients are required.
\subsection{UV insensitivity of winding corrections}
\label{sec:IRwind}
\begin{figure}[t!]
 \begin{center}
 \hskip 4mm\includegraphics[width=3.2in]{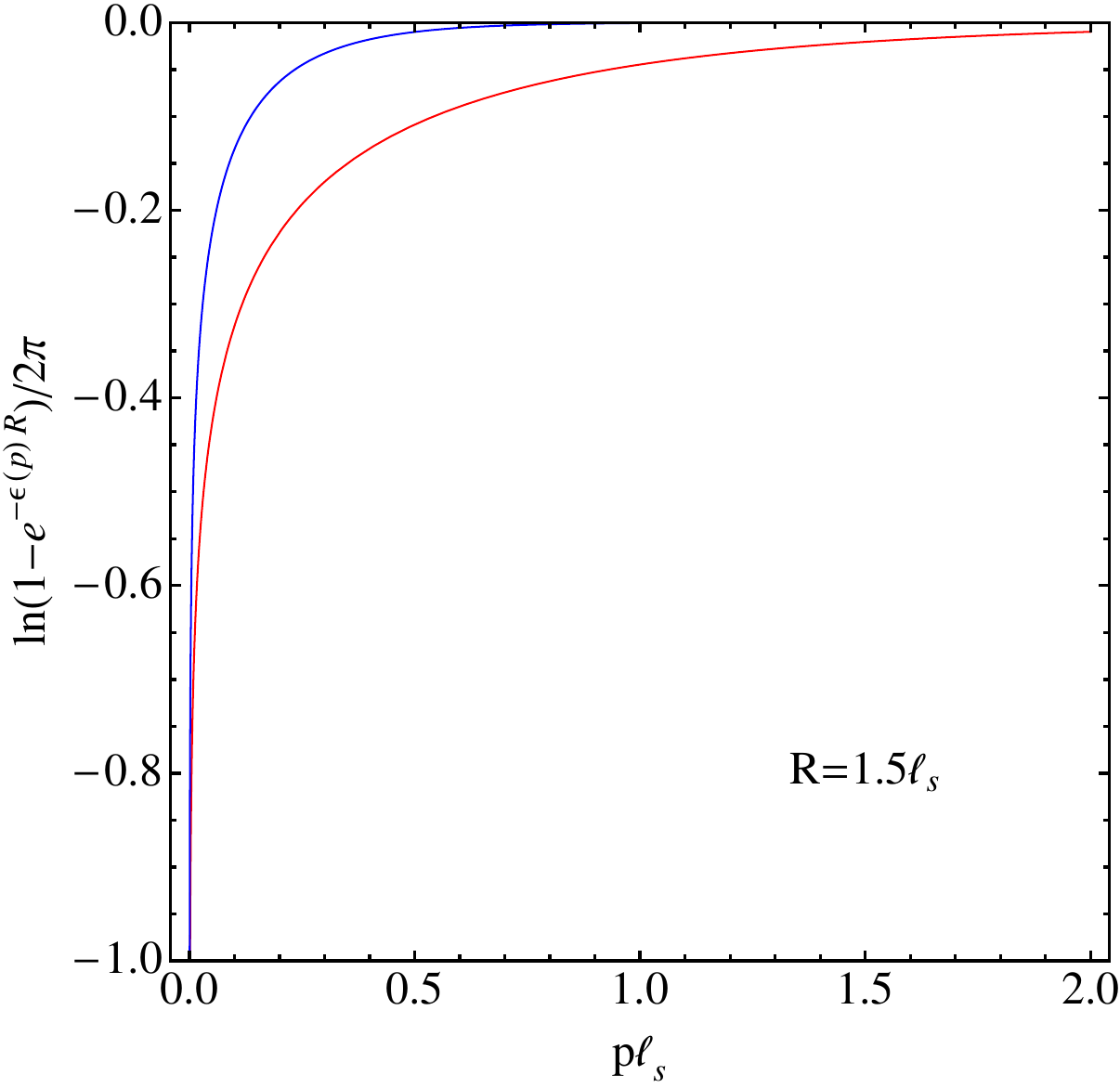}
 \caption{The integrand appearing in the winding corrections as a function of momentum. The blue line represents the integrand for the ground state. The red line shows the integrand for left-moving pseudo-particles in the presence of a right-moving excitation with one unit of KK momentum or vice-versa. The radius of the circle is taken to be $R=1.5\,\ell_s$, smaller than radii we are typicall interested in. Nevertheless, the integral is dominated by soft pseudo-particles and is rather insensitive to the UV behavior of the scattering amplitude. }
 \label{fig:thermal}
 \end{center}
\end{figure}

It is apparent from the above discussions that winding corrections are more subtle and harder to account for than the ABA part of the finite volume spectrum. 
In particular, if one is interested in a specific state with a fixed number of particles, solving the ABA requires diagonalization of the $S$-matrix only in that sector.
One the other hand, accounting for winding corrections always involves a complete diagonalization of the $S$-matrix for an arbitrary number of particles.
At some point later on (when discussing the resonance contribution in section~\ref{sec:resonance})  we will find ourselves in the situation, when the ABA part is straightforward to write down and solve. At the same time a complete diagonalization of the $S$-matrix is currently unavailable, and the winding corrections cannot be accounted for.

In a situation like this in massive theories it is a common practice to neglect the winding corrections, given that these are now exponentially suppressed.
In section~\ref{sec:resonance} we will employ a similar strategy and use lower order approximation to the phase shift in the winding part of the TBA than in the ABA part. Heuristically this may be justified by noting that the problem arises due to massive resonant contribution, and the same justification as in the massive case applies.
In fact, there is a general reason for winding corrections to be less sensitive to the UV physics than the asymptotic part.
 The integrals over the thermal bath in the full TBA system are exponentially cut off for momenta above $q \sim 1/R$, which is smaller than the characteristic momenta of real particles. In the free theory approximation the latter are of order $2 \pi/R$. We illustrate this point for the GGRT theory in Fig.~\ref{fig:thermal}. 
 We see, that even for a very short compactification radius, the winding integral is dominated by rather soft momenta and, as a consequence, is not very sensitive to higher order corrections to the amplitude.

\section{Energy Levels of Flux Tubes}
\label{sec:levels}
We are now in a good position to move on to the main topic of the paper and to apply the TBA technique to the calculation of the flux tube spectra.
The first step is to calculate the world sheet  $S$-matrix perturbatively in the $p\ell_s$-expansion. As explained in section~\ref{GGRT}, the phonon scattering on the world sheet of a flux tube is universal up to $\ell_s^4$. The corresponding amplitudes were calculated in \cite{Dubovsky:2012sh}. At this order there is no particle production, so the $S$-matrix is integrable and completely determined by the two particle elastic amplitudes. To describe the latter it is convenient to characterize two-particle states according to their quantum numbers under the unbroken group $O(2)$ of rotations in the transverse plane. One finds one scalar $|s\rangle$, one pseudoscalar $|p\rangle$ and two components $|t,\pm\rangle$ of the symmetric tensor  $O(2)$ representations. 
Introducing creation operators for  states of definite helicity, {\it i.e.}, eigenstates of the continuous $SO(2)$ rotations in the transverse $(X^2,X^3)$ plane,
\be
\label{helicitya}
a^\dagger_{l(r)\pm}=a^\dagger_{l(r)2}\pm i a^\dagger_{l(r)3}\;,
\ee
the corresponding states take the form
\begin{gather}
|s\rangle=(a^\dagger_{l+}a^\dagger_{r-}+a^\dagger_{l-}a^\dagger_{r+})|0\rangle
\nonumber\\
|p\rangle=(a^\dagger_{l+}a^\dagger_{r-}-a^\dagger_{l-}a^\dagger_{r+})|0\rangle
\nonumber
\\
|t,\pm\rangle=a^\dagger_{l\pm}a^\dagger_{r\pm}|0\rangle\;.
\label{spt}
\end{gather}
The two-particle $S$-matrix is diagonal in the basis (\ref{spt}) and at order $\ell_s^4$ reduces to the elastic scattering phases in each of the channels, which are equal to
\begin{gather}
\delta_{s(p)}=\delta_{GGRT}+\delta_{PS}+{\cal O}(\ell_s^6)
\label{spphase}
\\
\delta_{t}=\delta_{GGRT}-\delta_{PS}+{\cal O}(\ell_s^6)
\label{tphase}\;, 
\end{gather}
where $\delta_{GGRT}$ is the GGRT phase shift (\ref{GGRTphase}), and $\delta_{GGRT}$ is the Polchinski--Strominger phase shift given by
\be
\label{PSphase}
2\delta_{PS}=\frac{26-D }{24\pi} \ell_s^4 (p_l p_r)^2\;,
\ee
where we restored the dependence on the dimension $D$ of the target space-time\footnote{Of course, for $D\neq 4$ the expressions (\ref{spt}) should be modified, and the pseudo-scalar representation turns into an antisymmetric tensor. }.  The appearance of the critical string dimension $D_c=26$ in 
the PS phase shift (\ref{PSphase}) indicates that it introduces qualitatively new effects as compared to the leading GGRT phase shift. Indeed, one can show that the PS phase shift is responsible for the eventual breaking of integrability  on the world sheet of a non-critical string at higher order in the $\ell_s$ expansion.

At order $\ell_s^4$,  which we are working in, the theory is still integrable, but is not reflectionless anymore. The PS shift removes the degeneracy between phase shifts in different channels; the phase shift in the tensor channel is different from the one in the scalar and pseudoscalar channels.
As a consequence, annihilation transitions like $a^\dagger_{l2}a^\dagger_{r2}|0\rangle\to a^\dagger_{l3}a^\dagger_{r3}|0\rangle$ are possible at this order. 

As a result, in general, one expects that the reflectionless TBA, described in section~\ref{sec:diagrams}, can no longer be applied. For general $D$, this is indeed the case. However, $D=4$, when the string only two transverse directions, is special. Switching to the helicity basis (\ref{helicitya}),  allows to diagonalize the $S$-matrix for an arbitrary number of particles. Hence, for two flavors, we can still apply the full reflectionless excited TBA system described in section~\ref{sec:diagrams}. The only modification is that the TBA particles have to be labeled by their helicities
rather than by $O(2)$ flavors. The corresponding phase shifts are given by
\begin{eqnarray}
&&\delta_{++}=\delta_{--}=\delta_t\;,\nonumber\\
&&\delta_{+-}=\delta_{-+}=\delta_{s(p)}\;.
\label{pmphases}
\end{eqnarray}
%To avoid confusion, note that the presence or absence of annihilation is, of course, an invariant physical property independent of the choice of the field basis.
%In the helicity basis, where the $S$-matrix is diagonal, it manifests itself through the presence of even in $s$ terms in the phase shifts (\ref{pmphases}). These
%are forbidden by real analyticity for the real field basis, where the $S$-matrix is instead non-diagonal.

Before concluding the section, let us briefly comment on the $D=3$ case since we will discuss the $D=3$ lattice data. In that case one finds a single two-particle state with zero total momentum. The PS amplitude in this case vanishes for kinematic reasons, and the world sheet $S$-matrix agrees with the GGRT $S$-matrix at order $\ell_s^4$.

Let us now apply the TBA approach to various states (and theories).

\subsection{Ground state energy}
As discussed in section~\ref{GGRT}, the ground state is the only state for which the conventional $\ell_s/R$-expansion is adequate for explaining the data.
 The vacuum matrix element of the PS-operator (\ref{PSoperator}) vanishes. So the ground state energy deviates from that in the GGRT model only at order $(\ell_s/R)^7$. As shown in Fig.~\ref{fig:ground} the sum of the universal terms agrees very well with the lattice data.
One finds equally good agreement by applying the TBA method. Using the leading $\ell_s^2$-order expression for the phase shift ({\it i.e.}, the GGRT phase shift), the solution of the TBA equations with $N=\tilde{N}=0$ reproduces the GGRT vacuum energy (see \cite{Dubovsky:2012wk} for details). Figure~\ref{fig:ground} shows that the two result are undistinguishable at the currently available level of precision of the lattice data.
 
 Including the PS phase shift does not change the answer, in agreement with the result from the $\ell_s/R$ expansion.
 Indeed, in this case all TBA particles are characterized by a single pseudo-energy $\epsilon(q)$, which is obtained by solving a single TBA constraint which
takes the form (c.f. with a general form of the TBA constraints (\ref{TBAepsl}),  (\ref{TBAepsr})),
\be
\label{vacuumTBA}
\epsilon(q)=q+{1\over 2\pi R}
\int dq'\left(
\frac{d\,2\delta_{++}(q,q')}{dq'}+
\frac{d\,2\delta_{+-}(q,q')}{dq'}\right)\ln\left(1-e^{-R\eps (q')}\right)\;.
\ee
The PS-contribution cancels in the sum of the phase shifts and one obtains exactly the same pseudo-energy as in the GGRT theory, and correspondingly the same result for the vacuum energy.

\subsection{Purely left(right)-moving states}
\begin{figure}[t]
 \begin{center}
 \includegraphics[width=3.35in]{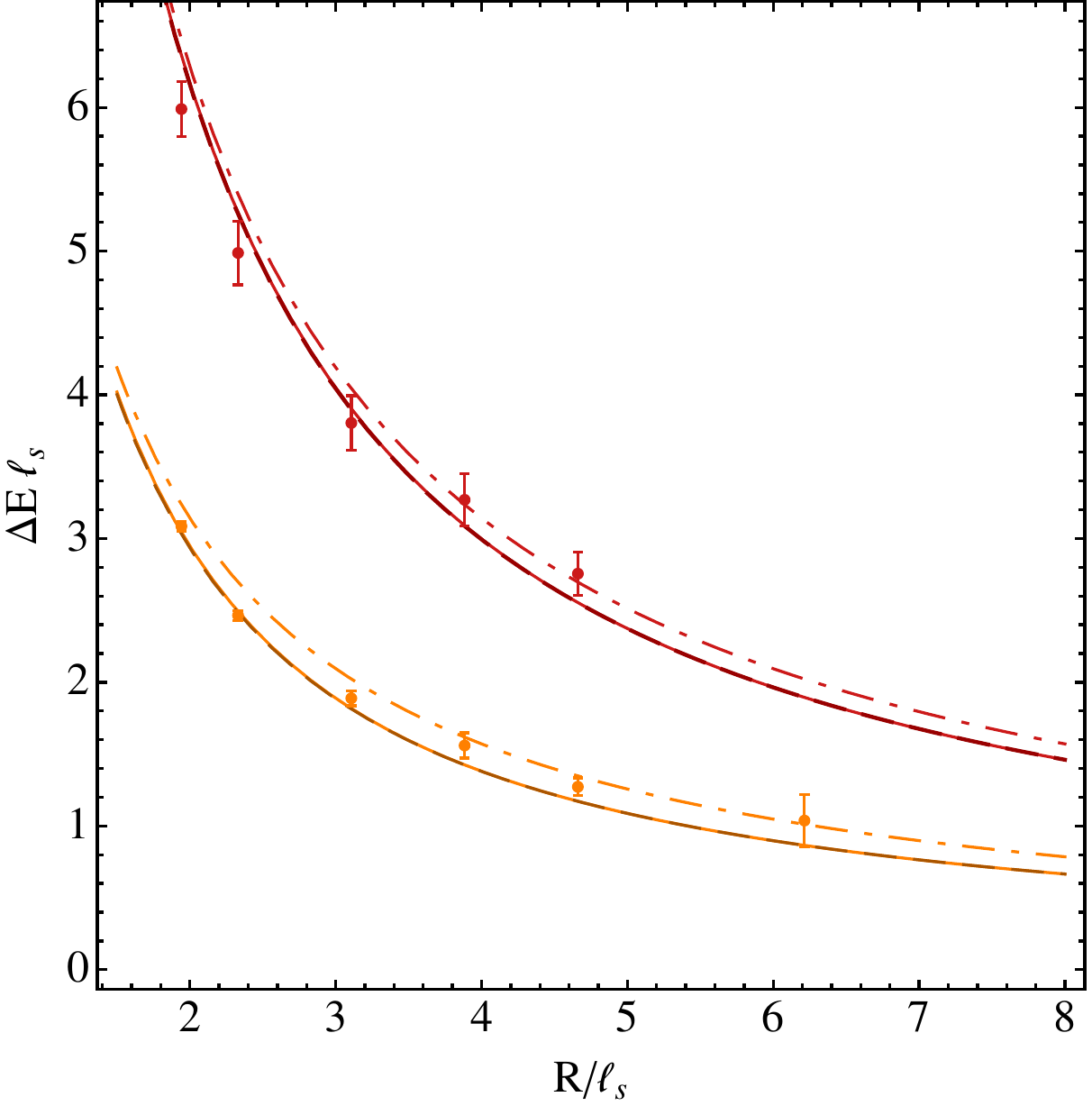} 
 \caption{This plot shows $\Delta E=E-R/\ell_s^2$ as a function of the length of the flux tube for the states with left-movers with one and two units of KK momentum in orange and red, respectively. The data is taken from~\cite{Athenodorou:2010cs}. The solid lines show the theoretical predictions derived from equations (\ref{TBAc})-(\ref{TBAe}) with the PS interaction taken into account to all orders. The darker dashed lines show the result in which only the GGRT phase is included. The dot-dashed line shows the ABA or in this case equivalently the free theory result.}
 \label{fig:oneKKTBA}
 \end{center}
\end{figure}

Let us turn to states which contain only left- (or right) moving real phonons, {\it i.e.} \mbox{$\tilde N=0$} and arbitrary $N$.  This is the simplest class of states for which the standard $\ell_s/R$-expansion breaks down even for relatively long strings as can be seen Fig.~\ref{fig:oneKK}. Fortunately, these states are still simple from the point of view of the TBA. The asymptotic Bethe Ansatz is especially simple, because there are no interactions between left-movers. Accounting for windings by keeping the leading GGRT part of the scattering amplitude, one obtains the GGRT expression as an approximation for the energies of these states. As we already discussed, this approximation works very well.

To find the result for the amplitude to order $\ell_s^4$ given in equation (\ref{pmphases}), we have to solve the TBA constraints (\ref{TBAepsl}) and  (\ref{TBAepsr}) for four different pseudo-energies $\eps^\pm_l$, $\epsilon^\pm_r$. 
As a consequence, different from the ground state, the energies acquire a dependence on the PS phase shift (even though the PS operator
has zero matrix elements for these states, so there is no $(\ell_s/R)^5$-correction in the standard perturbative expansion). 
The TBA equations together with the explicit expressions for the phase shifts (\ref{pmphases}) imply that pseudo-energies are now complex and of the form
\be
\label{quadep}
\eps_{l(r)}^\pm(q)=c_{l(r)}q\pm i d_{l(r)}q^2
\ee
with real $c_{l(r)}$ and $d_{l(r)}$.
% The dependence of the pseudo-energies on the momenta follows from the TBA constraints (\ref{TBAepsl}), (\ref{TBAepsr}) after substitution of the explicit expressions for the phase shifts (\ref{pmphases}). It is straightforward to check that the reality conditions for $c$'s and $d$'s are compatible with all the TBA equations. One can also arrive to them by solving the TBA equations perturbatively in the PS phase shifts. In this approach one starts with the GGRT solution, which corresponds to real $A$'s and zero $B$'s. By perturbing around this solution one always preserves the reality of $A$'s and $B$'s.

The resulting set of equations for the coefficients $c$ and $d$ are straightforward to solve numerically. The result is presented in Fig.~\ref{fig:oneKKTBA}. The figure shows both the result in which the windings are evaluated for the phase shift at order $\ell_s^4$ as discussed here and for the GGRT phase.
In accordance with our earlier discussion about the UV insensitivity of the winding corrections the effect of the PS phase is very small ($\lesssim 0.5\%$).

The GGRT winding corrections are in fact also small. This can be seen in  Fig.~\ref{fig:oneKKTBA} which also shows the ABA result, or equivalently the free theory answer, for the energies.
The physics of these states is very simple. To a very good approximation they are just collections of free phonons. 

\subsection{States with left- and right-mover and a new massive state}
\label{sec:resonance}
We now consider the states with one left- and one right-moving particle each carrying one unit of momentum, {\it i.e.}, $N=\tilde{N}=1$. These are the lowest energy states for which the ABA is non-trivial and we will finally be able to see all ingredients of the TBA method at work. Figure~\ref{fig:twoKK} shows that for these states the naive derivative expansion does not provide a good approximation for strings with lengths accessible on the lattice.

\begin{figure}[h]
 \begin{center}
 \includegraphics[width=3.35in]{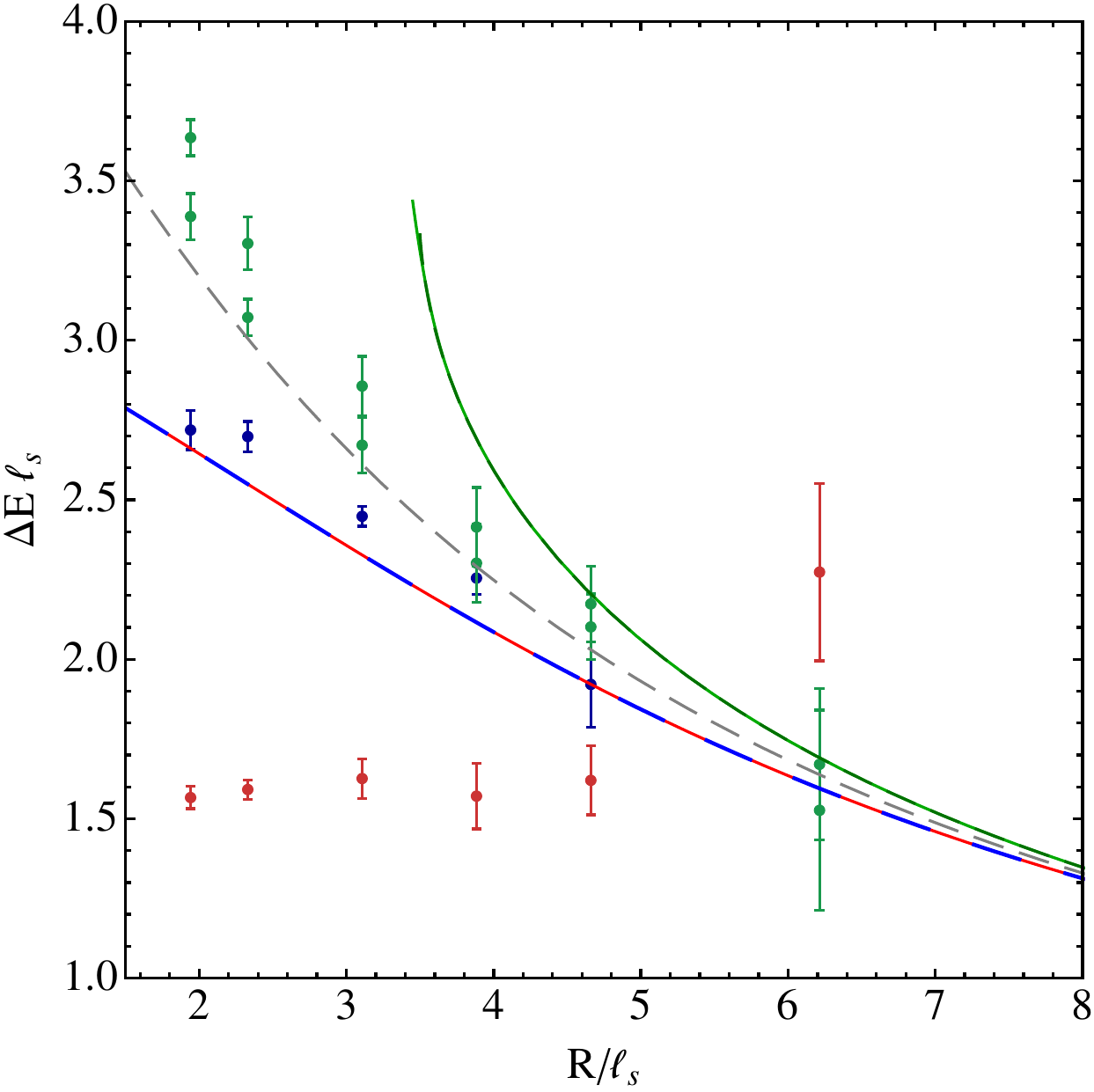} 
 \caption{This plot shows $\Delta E=E-R/\ell_s^2$ as a function of the length of the flux tube for the lowest lying states containing both left- and right-movers. The parity even and parity odd states with spin 0 are shown in blue and red, respectively. The states with spin 2 are shown in green. The data is taken from~\cite{Athenodorou:2010cs}. The red, blue, and lighter green lines show the theoretical predictions derived from equations (\ref{TBAc})-(\ref{TBAe}) with the PS interaction included to all orders. The darker green dashed line shows the result when both the GGRT and PS phase is taken into account in the asymptotic Bethe Ansatz, but but only the GGRT phase is taken into account for winding corrections. The two are virtually indistinguishable, once again showing the UV insensitivity of the winding corrections. The dashed gray line shows the result for the GGRT theory.}
 \label{fig:e11nores}
 \end{center}
\end{figure}

As before, keeping the GGRT part of the phase shift in the TBA system results in the GGRT expression for the energies.  From Fig.~\ref{fig:twoKK} one finds that it provides a reasonable approximation for the scalar and tensor levels, but not for the pseudoscalar.  Let us now include the PS phase shift. Similarly to  purely left-moving states we are after four complex pseudo-energies $\eps^\pm_l$, $\epsilon^\pm_r$. They are again of the quadratic form (\ref{quadep}) and satisfy the same reality conditions. Since our states satisfy $p_l=p_r$, the TBA system imposes the additional relation $\epsilon_l^\pm(q)=\epsilon_r^\pm(q)$ for the scalar and pseudo-scalar state and $\epsilon_l^\pm(q)=\epsilon_r^\mp(q)$ for the tensor states. The resulting equations for $c$ and $d$ are again readily solved numerically.
The results are shown in Figure~\ref{fig:e11nores}.
The convergence of the TBA result is significantly better than that of the $\ell_s/R$-expansion. For the scalar and tensor levels we also find significantly improved agreement with the lattice data. This is noteworthy especially because so far we have not introduced any free parameter in our analysis in addition to $\ell_s$, which is fixed from the ground state data, just like in the $\ell_s/R$-expansion. So the curves presented in Fig.~\ref{fig:e11nores} are the results of a calculation from first principles.

 The  improvements in the convergence of the perturbative expansion are more prominent for the scalar state than for the tensor states.
The reason for this is that in the TBA method the perturbative approximation enters in the calculation of the scattering amplitudes.
How good the perturbative expansion is, is controlled by how soft the phonon momenta $p\ell_s$, that comprise the states. These momenta are determined from solving the TBA system and take different values in the different channels for the same value of $R$.
The PS correction in the (pseudo)scalar channels adds to the tree-level phase shift. In the tensor channel it has opposite sign so that the phase shift grows more slowly. 
In agreement with the discussion at the end of section~\ref{sec:ABA},  the phonon momenta are then softer in the (pseudo)scalar sectors so that the perturbative expansion behaves better.

To demonstrate this effect, the theoretical curves on the plot are terminated when the momenta of the particles become large enough so that the one-loop contribution to the phase shift  $\delta_{PS}$ becomes equal to the tree level one $\delta_{GGRT}$. This happens when $p\approx\, 1.8\ell_s$. 

Even though the PS contribution to the phase shift affects these states significantly, its effect on the winding corrections is still negligible because the winding corrections are UV insensitive as shown earlier. To illustrate this explicitly for these states, we also solved the TBA system by including the PS phase shift in the asymptotic Bethe Ansatz but neglecting it in all winding contributions ({\it i.e.}, in the TBA constraints (\ref{TBAepsl}), (\ref{TBAepsr}) and in the integral terms in the momenta quantization conditions (\ref{TBApl}), (\ref{TBApr})). The result is shown in Figure~\ref{fig:e11nores} together with the exact treatment. The difference is again less than $\sim 0.5\%$.

The improved theoretical control makes it manifest that the anomalous behavior of the pseudoscalar level is a genuinely new physical effect and is unrelated to the bad convergence of the expansion. At this order the scalar and pseudoscalar states, for which the expansion is well-behaved, are predicted to be degenerate. However, the observed splitting between the scalar and pseudoscalar states is larger than the splitting (both predicted and observed) between the scalar and tensor states even for relatively long strings. It is then implausible to expect that this discrepancy will disappear by including  higher order contributions to the world sheet $S$-matrix.

This strongly suggests that to explain the anomalous behavior of the pseudoscalar level we need to reconsider the basic assumptions underlying our calculation and add a qualitatively new input. An important hint suggesting the missing ingredient comes from observing that the energy of the pseudoscalar level is practically independent from the length of a flux tube. This suggests that we are observing a light massive excitation on the world sheet of a flux tube -- a new particle. A similar explanation for the energy of the pseudoscalar level was suggested earlier in \cite{Athenodorou:2010cs}.

%We see that the scalar and tensor levels do fit the data, however, the lowest pseudo scalar level is significantly lower than predicted by the theory. Moreover, the very splitting between scalar and pseudo scalar levels is not captured by our theory. The important observation is that dependence of the energy of this level on the radius is very weak - it looks a lot like a massive particle. This suggests that we are observing the lightest massive excitation on the world sheet of the flux tube - a new particle. This explanation for the energy of the pseudo scalar level was earlier suggested earlier in \cite{Athenodorou:2010cs}.

It is straightforward to incorporate such a state in our effective string theory framework. The minimal possibility is to introduce a new massive pseudoscalar field $\phi$ on the flux tube world sheet. At the leading  order in the derivative expansion interactions of such a field with the Goldstones are described by the following Lagrangian,
%\be
%\mathcal{L}_{\phi} =-M^2 \phi^2 -(\partial \phi)^2
%\ee
\be
 \label{Lphi}
 \mathcal{L}_{\phi} =-\half (\partial \phi)^2-\half m^2 \phi^2 -  {\alpha\over  8\pi } \phi \eps^{ij} \eps^{\alpha \beta} \partial_{\alpha} \partial_{\gamma}X^i\partial_{\beta} \partial^{\gamma} X^j+\dots\;,
\ee
where dots stand for terms, which are higher order in fields and derivatives. In particular, these include model independent quartic $\phi\phi XX$ couplings originating from the covariant completion of the kinetic and mass term for $\phi$. 

The presence of four-derivatives in the leading pseudoscalar $\phi XX$ coupling in (\ref{Lphi}) is dictated by non-linearly realized Lorentz invariance.
It requires that every term in the action corresponds to the expansion of some geometric invariant (see, e.g., \cite{Cooper:2013kga} for a recent discussion). The invariant that corresponds to the interaction term in (\ref{Lphi}) is rather special and deserves some attention. It originates from  
\be
\label{axionint}
 \nu={\alpha\over  8\pi } \phi K^i_{\alpha\gamma}K^{j\gamma}_\beta\epsilon^{\alpha\beta}\epsilon_{ij}\,,
\ee
where $K^i_{\alpha\gamma}$ is the extrinsic curvature of the world sheet. Thus, $\phi$ is coupled to the topological invariant known as the self-intersection number of the string world sheet. The existence of this world sheet $\theta$-term for a string in a four-dimensional target space was pointed out by Polyakov \cite{Polyakov:1986cs}, and it was suggested that it should be generated on the flux tube world sheet in the presence of the bulk $\theta$-term \cite{Mazur:1986nr}. Given this coupling, it is natural to refer to the field $\phi$ as the world sheet axion.

This axion is not a stable particle, so it should not be added to the set of asymptotic states in the TBA system. However, it does contribute to the scattering of Goldstones. In particular it appears as a resonance in the pseudoscalar channel, where its effect is most pronounced. A diagrammatic calculation using the action~\eqref{Lphi} to leading order in $\alpha$ gives the following contribution to the two-particle phase shift
 \be
\label{delta_res}
2\delta_{res}(p)=\sigma_1{\alpha^2\ell_s^4{p}^6\over 8\pi^2(4{p}^2+m^2)}+2\sigma_2 \tan^{-1}\l{\alpha^2\ell_s^4{p}^6\over 8\pi^2 (m^2-4{p}^2)}\r\,.
\ee
with $\sigma_1=(-1,1,1)$, $\sigma_2=(0,0,1)$,
for scalar, symmetric, and pseudoscalar channels, respectively. The $\sigma_2$-term represents the resonant $s$-channel contribution, while the $\sigma_1$-term
arises from the $t$- and $u$-channels.

Accounting for the pseudoscalar resonance in the winding contributions is problematic because switching to the helicity field basis (\ref{helicitya}) no longer diagonalizes the full $S$-matrix. Already in the two-particle sector the phase shifts (\ref{delta_res}) now take different values in the scalar and pseudoscalar channels 
(which is, of course, the reason we introduced the resonance in the first place). As a consequence, we can no longer include the PS contribution into winding corrections either. However, we have already seen that the winding corrections are not UV sensitive and that the error introduced by not including the PS contribution in the winding corrections is negligible ($\lesssim 0.5\%$). From now on, we will thus account for the full phase shifts only in the ABA part of the generalized momentum quantization conditions  (\ref{TBApl}), (\ref{TBApr}) and everywhere else keep only the GGRT contribution. This significantly simplifies the TBA system. The pseudo-energies become real,  independent of the flavor of the particles and linear in
momenta,
 \[
 \eps^1_{l(r)}(q)=\eps^2_{l(r)}(q)=c q\;.
 \] 
This converts the TBA equations into the following simple system of algebraic equations
\be
\label{TBAc}
c=1+\frac{p\ell_s^2}{R}-\frac{\pi(D-2)}{12 R^2 c}\ell_s^2,% \quad c_r=1+\frac{p_{li}\ell_s^2}{R}-\frac{\pi(D-2)}{12 R^2 c_l}\ell_s^2,
\ee
\be
\label{TBAp}
pR+ 2\delta(p)-\frac{\pi(D-2)}{12Rc}\ell_s^2p=2 \pi N,% \quad p_{ri}R+\sum\limits_j 2\delta(p_{lj},p_{ri})-\frac{\pi(D-2)}{12Rc_l}\ell_s^2p_{ri}=2 \pi \tilde{N}_i
\ee
where $N=1$, and the expression for the energy is
\be
\label{TBAe}
\Delta E=2p-\frac{\pi(D-2)}{6 R c}\;.
\ee
Depending on the state, the phase shift in equation~\eqref{TBAp} is given by the sum of one of (\ref{spphase}), (\ref{tphase}) and of (\ref{delta_res}).

\begin{figure}[t]
 \begin{center}
 \includegraphics[width=3.35in]{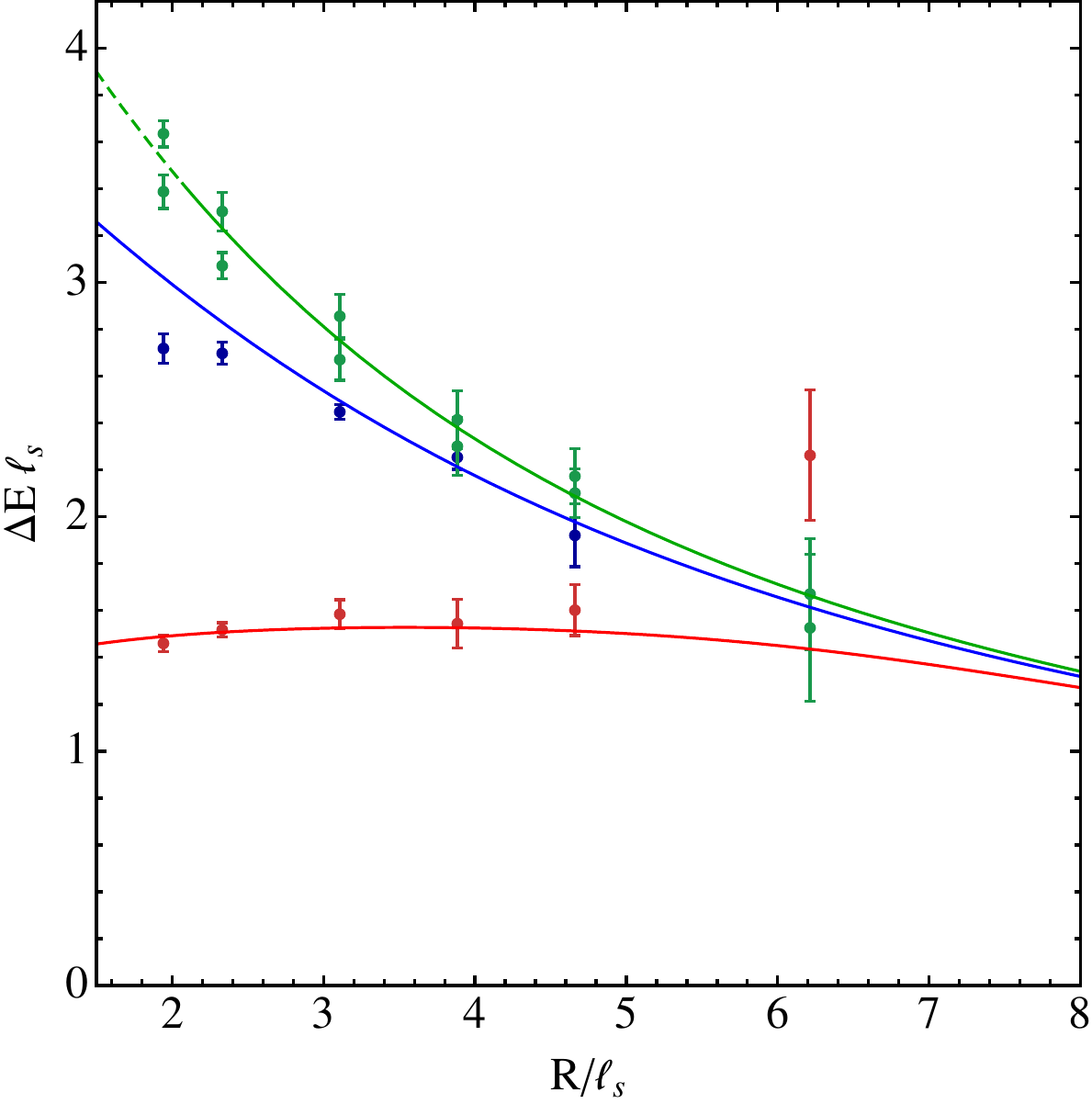} 
 \caption{This plot shows $\Delta E=E-R/\ell_s^2$ as a function of the length of the flux tube for the lowest lying states containing both left- and right-movers. The data is again taken from~\cite{Athenodorou:2010cs} and the coloring is as in Figure~\ref{fig:e11nores}. The red, blue, and green lines show the theoretical predictions derived from equations (\ref{TBAc})-(\ref{TBAe}) with GGRT, PS and resonance contribution to the phase shift included in the asymptotic Bethe Ansatz, but with winding corrections only taken into account for the GGRT contribution. Lines are shown as dashed where the PS contribution becomes larger than the GGRT contribution.}
 \label{fig:e11tba}
 \end{center}
\end{figure}

The axion introduces two free parameters, the mass $m$ and the coupling $\alpha$ (or, equivalently, the width).
We determine them by fitting the model to the data and find 
\be
m\ell_s=1.85^{+0.02}_{-0.03}\; ,\;\;\alpha\ell_s^{-2}=9.6\pm 0.1\;.
\ee
In physical units this corresponds to approximately 750 MeV, which is about a half of the mass of the lightest glueball.
It should be kept in mind that the presented error bars reflect the statistical uncertainty only. We estimate the systematic errors to be comparable.
The results are presented on the Fig.~\ref{fig:e11tba}. The lines on the plot become dashed where $\delta_{PS}$ becomes equal to $\delta_{NG}$.
We see that including the axion not only provided a very good fit for the pseudo-scalar state, but also significantly improved the fit in other channels thanks to the $\sigma_1$-term in (\ref{delta_res}). Note that the sign of this correction cannot be changed by varying the parameters so that this is rather non-trivial.
The best fit values correspond to a relatively narrow resonance with a width equal to 
\[
\Gamma=0.39/\ell_s=0.21\,m\,.
\]

\subsection{Determination of  phase shifts from the data and excited levels}
Another advantage of the method developed in this paper is that it allows us to present the data in a new way. Similar to the standard procedure used to extract scattering amplitudes from  lattice calculations \cite{Luscher:1986pf}, we can use the system of equations (\ref{TBAc})-(\ref{TBAe}) to solve for $p$ and $\delta$ given $\Delta E(R)$. The only difference is that we include winding corrections because our phonons are massless. This alternative way of presenting data allows us to directly visualize the presence of a resonance and the extent to which the resonance improves the fit in the scalar and tensor channel. In addition, it has the advantage that we can combine different excited states in the same plot because they probe the same underlying scattering amplitudes. As an example, let us look at the phase shift for the states with one left- and one right-mover as a function of the center of mass energy extracted from the data for the energy levels. In this case the solution can be written in a relatively compact form
\be
%p_l=p_r=\frac{3 \Delta E^2\ell_s^2 + 2 \pi + 6 \Delta E R}{6 (\Delta E \ell_s^2+ 2 R)}
p_l=p_r=\frac{\Delta E}{2}+\frac{ \pi }{6 (\Delta E \ell_s^2+ 2 R)}\,,
\ee
\be
%2\delta=\frac{42\pi \Delta E^2 \ell_s^2 + 8 \pi^2 - 9 \Delta E^3 R \ell_s^2+ 144\pi \Delta E R - 36 \Delta E^2 R^2 + 
% 120 \pi R^2 /\ell_s^2- 36 \Delta E R^3/\ell_s^2}{18 (\Delta E \ell_s^2 + 2 R)^2}
2\delta=2\pi-\frac{\Delta E R}{2}+\frac{\pi}{18} \frac{3\Delta E^2 \ell_s^4 +2 \pi \ell_s^2 -12 R^2}{ (\Delta E \ell_s^2 + 2 R)^2}\,.
\ee
\begin{figure}[t]
 \begin{center}
 \includegraphics[trim=4.5cm .5cm 5cm 1.cm,width=2.4in]{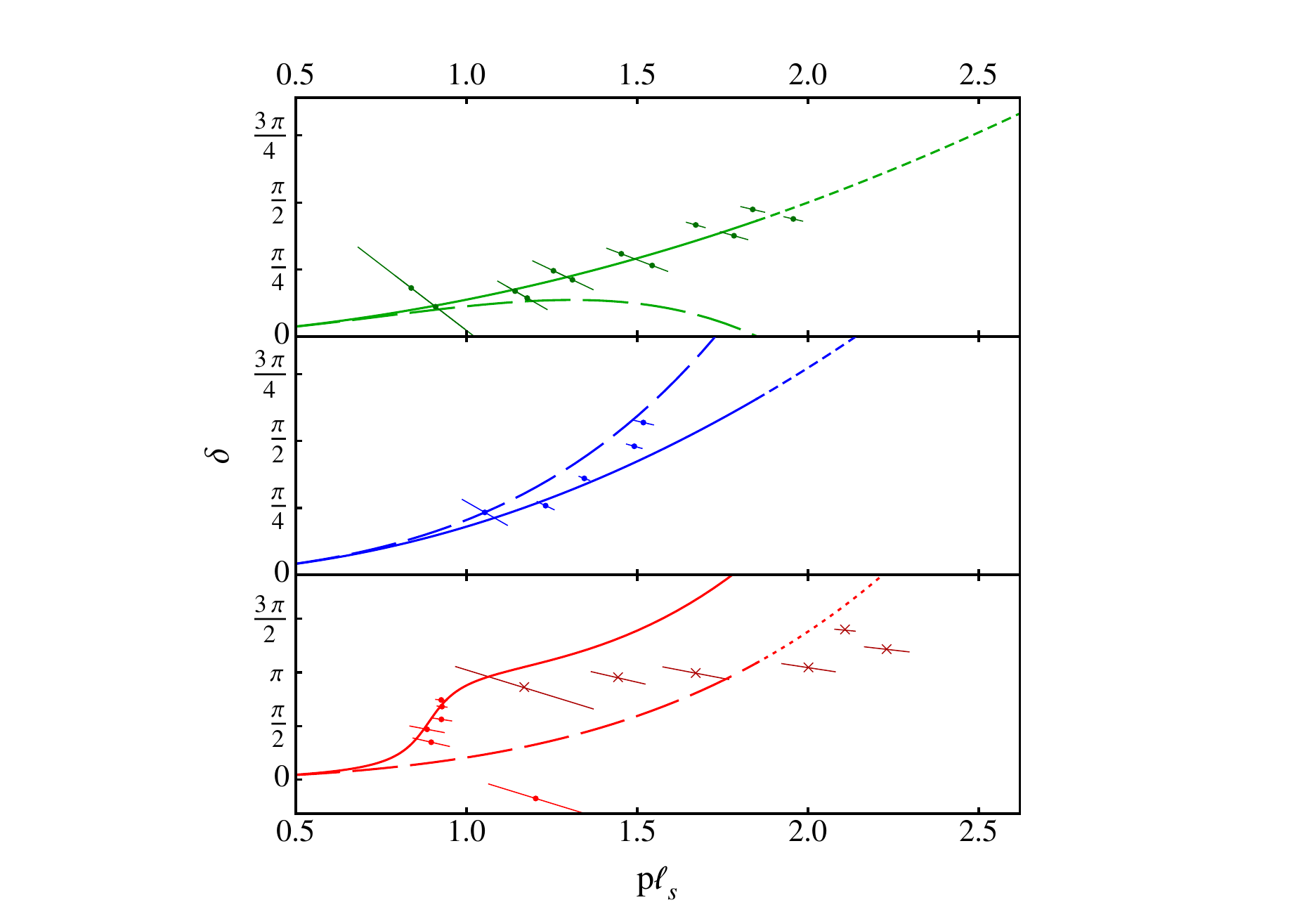} 
 \caption{This plot shows the scattering phase shift $\delta$ for two Goldstone bosons as a function of the center of mass momentum in the symmetric traceless, scalar, and antisymmetric channel in the top, middle, and bottom panel, respectively. The solid and the long dashed lines show the theoretical prediction with and without the world sheet axion, respectively.}
 \label{fig:resonance}
 \end{center}
\end{figure} 
\begin{figure}[t]
 \begin{center}
 \includegraphics[trim=4.5cm .5cm 5cm 1.0cm,width=2.4in]{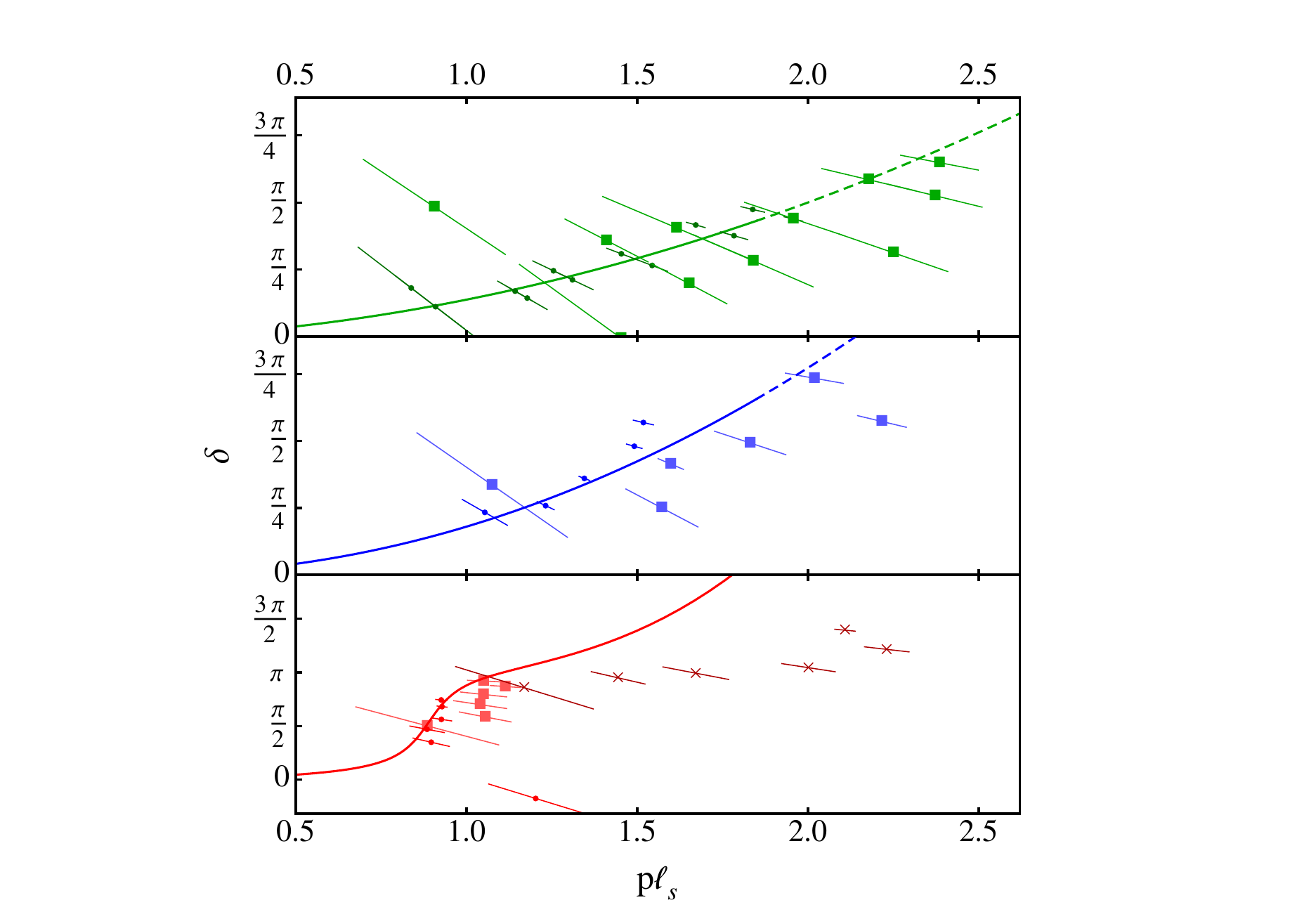} 
 \caption{This plot shows the scattering phase shift $\delta$ for two Goldstone bosons as a function of the center of mass momentum in the symmetric traceless, scalar, and antisymmetric channel in the top, middle, and bottom panel, respectively. The darker points show the lowest-lying states with zero total momentum, the crosses show the first excited pseudoscalar state, and the squares show the lowest-lying states with one unit of total momentum. The lines show the theoretical prediction.}
 \label{fig:resKK}
 \end{center}
\end{figure}

The resulting phase shift as a function of momentum extracted from the data is shown in Fig. \ref{fig:resonance} along with the theoretical predictions for  scattering phase shifts in  various channels. We also included the data for the next excited pseudoscalar level in the lower panel. The theoretical prediction and the data still agree for the excited state at  low momenta, but the agreement becomes noticeably worse as the momentum increases. Nevertheless, one clearly sees the characteristic resonance shape with a mass $m\simeq 1.8 - 1.9\ell_s^{-1}$. The middle and upper panels of the same figure show the scalar and tensor channels, and it is clear that in these channels there is no sign of a resonance.  The dashed curves represent the theory prediction in the absence of the resonance and only depend on one parameter, the string width $\ell_s$. The lower panel clearly shows that a new massive pseudo-scalar particle has to be introduced to explain the data. 

Of course, the introduction of a new massive pseudo-scalar state makes additional predictions. As already mentioned, it also affects the scalar and tensor states and improves the agreement between theory and lattice data for them. In addition, we should be able to give momentum to this particle so that it makes definite predictions for excited states with non-zero total momentum for which data is also available. If we extract phase shifts from the data for excited states with one left- and one right-mover with unequal momenta, we expect to find a resonance there as well. We show the result for the state in which the left-mover has one and the right-mover has two units of KK momentum in Fig.~\ref{fig:resKK} together with the theoretical prediction and the data for the state with zero total momentum, which we discussed earlier. Similarly, it makes predictions for the scalar and tensor states with non-zero total momentum. The states with one unit of total momentum for scalars and tensors are also shown in Fig.~\ref{fig:resKK}. The phase shifts extracted from the different excited states agree relatively well, almost within the statistical errors of the lattice calculations. In particular, we do see the resonance not only in the state with total momentum zero but also in the state with non-zero momentum. The small discrepancy between the two can be attributed to two effects. We did not include the finite size corrections due to the resonance itself into our calculation. An estimate shows that this affects the data points corresponding to the shortest lengths more strongly (as one would expect), and bring the phase shifts from the state with total momentum zero and unity into slightly better agreement. The remaining difference seems to be due to discretization effects in the lattice calculations themselves that are also responsible for the splitting between the two tensor states.

\subsection{$D=3$ Yang--Mills}
These techniques can, of course, also be applied to existing lattice data in $D=3$ dimensions. In this case we have a single channel for two-particle scattering.
The non-linearly realized Lorentz invariance implies that the phase shift takes the form
\bea
2\delta=2\delta_{GGRT}+\mathcal{O}(\ell_s^6 s^3)\,.
\eea
The corrections here are non-universal. In particular, as already mentioned, the GGRT phase shift itself is compatible with non-linearly realized Lorentz symmetry for $D=3$.

\begin{figure}[t]
 \begin{center}
 \includegraphics[width=6in]{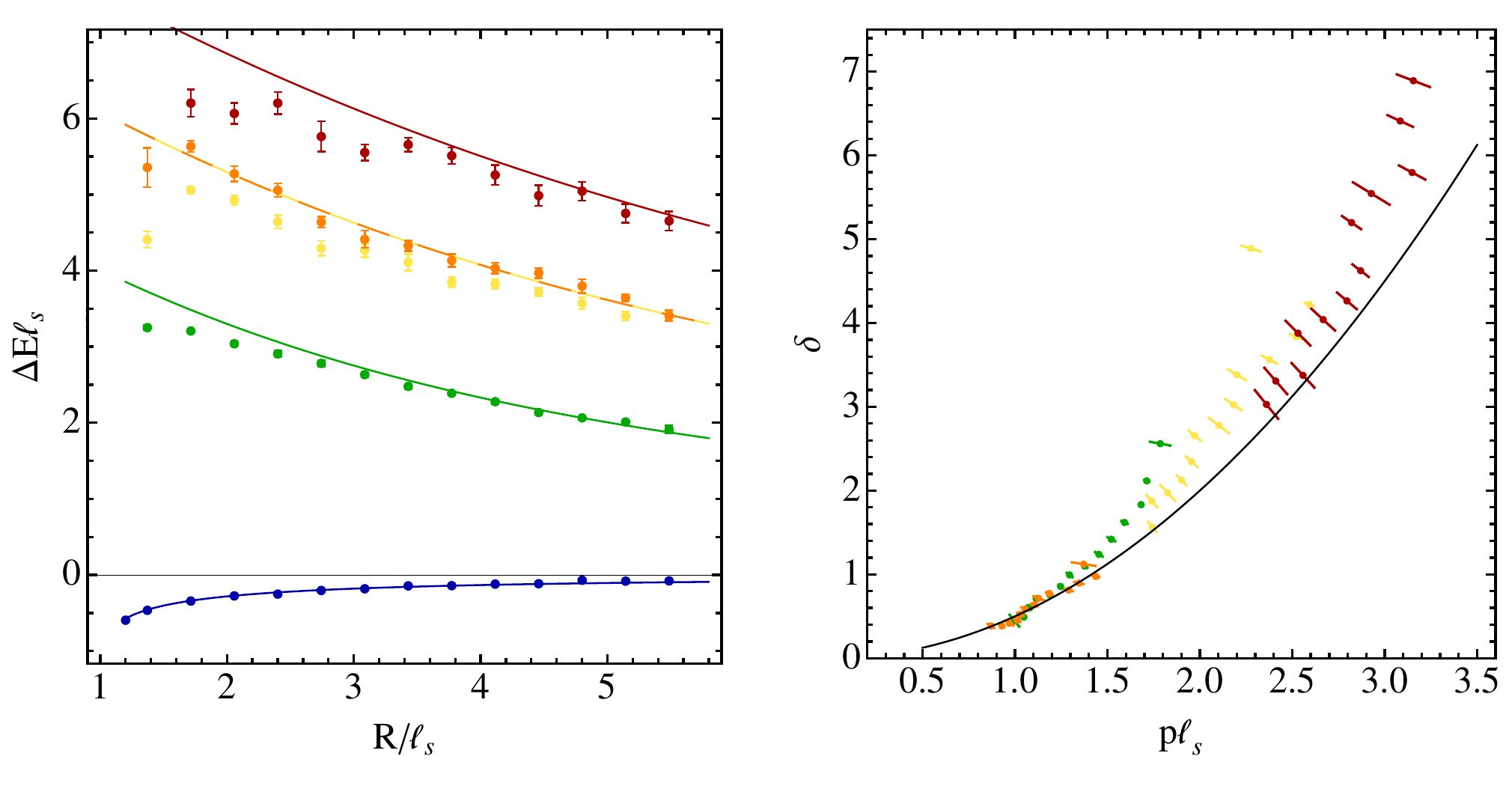} 
 \caption{This plot shows the energy and scattering phase shift $\delta$ for the lowest-lying parity even states with zero total momentum. The lines show the theoretical prediction of the GGRT theory.}
 \label{fig:E3d}
 \end{center}
\end{figure}

In this section, we will compare data from~\cite{Athenodorou:2011rx}, which is for gauge group SU(6) with $\beta=171$ and the Wilson loop in the fundamental representation, with the GGRT prediction. The result for the five lowest-lying states with an even number of phonons and zero total momentum is shown in Figure~\ref{fig:E3d}. We see that all states are in qualitative agreement with the GGRT phase shift and see no evidence for new light massive states. 
However, there are small quantitative differences between the GGRT prediction and the data. The energies of the states shown in yellow and orange correspond to states with two and four phonons and are predicted to be degenerate. However, they appear to be split in the data. Furthermore, the measured energies are systematically below the GGRT prediction. This suggests that the binding energy between the phonons is larger in the SU(6) gauge theory than in the GGRT theory which implies a phase shift that grows more rapidly, consistent with what is seen in the right panel of Figure~\ref{fig:E3d}. It is then natural to introduce corrections into the phase shift
\bea
2\delta=2\delta_{GGRT}+\gamma_3\ell_s^6 s^3\,,
\eea
and determine this leading correction from the data using the TBA, taking into account only the GGRT phase shift in the windings as before. Such a correction to the phase shift would follow from higher order geometric invariants in the Goldstone theory such as $R^2$ in the action, and we can trust our procedure provided the coefficient is small enough so that this is in fact a correction for the range of momenta of interest. Based on loop counting, one expects the coefficient to be of order $1/(2\pi)^2$, so that this should roughly be reliable for $p\ell_s\lesssim\sqrt{2\pi}$, which includes all data points of the first excited state for both two- and four-particle states, but only some of the second excited two-particle state. We extract $\gamma_3$ from the first excited two-particle state using the TBA equations~\eqref{TBAc}-\eqref{TBAe} as well as the first excited four-particle state using
\begin{figure}[t]
 \begin{center}
 \includegraphics[width=3.in]{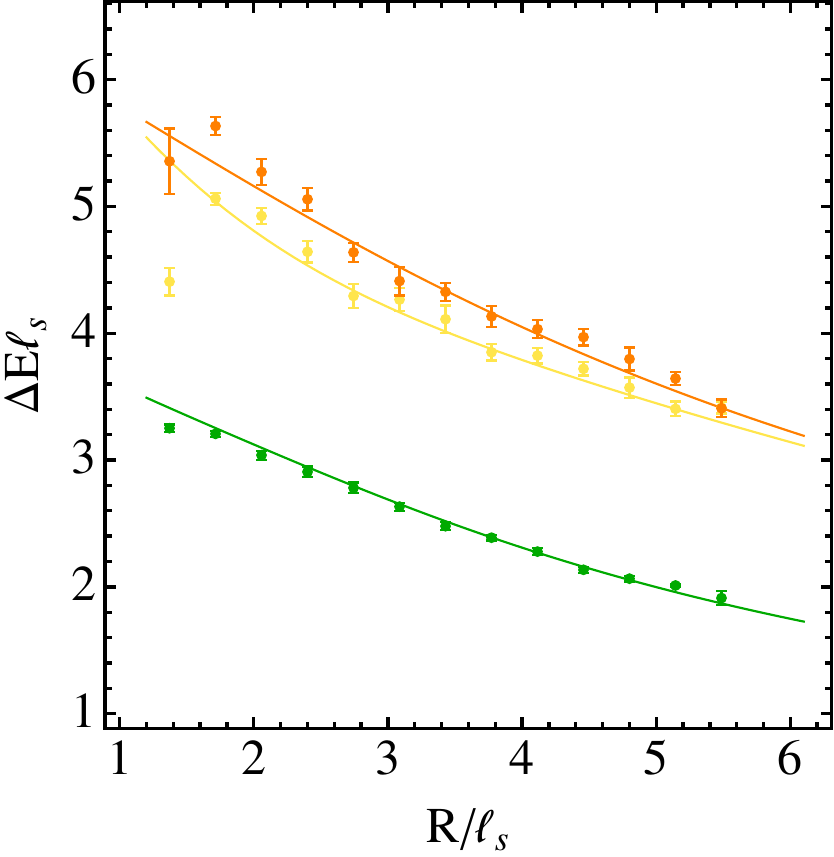} 
 \caption{This plot shows the energies of the states included in the fit for the correction to the scattering phase shift at order $\ell_s^6$. The lines show the theoretical prediction. For the second excited two-particle state, only data for six longest strings is included in the fit because the phonon momenta become too large.}
 \label{fig:E3dgamma}
 \end{center}
\end{figure}
\begin{figure}[h]
 \begin{center}
 \includegraphics[trim=1cm .8cm .1cm 1cm,width=6in]{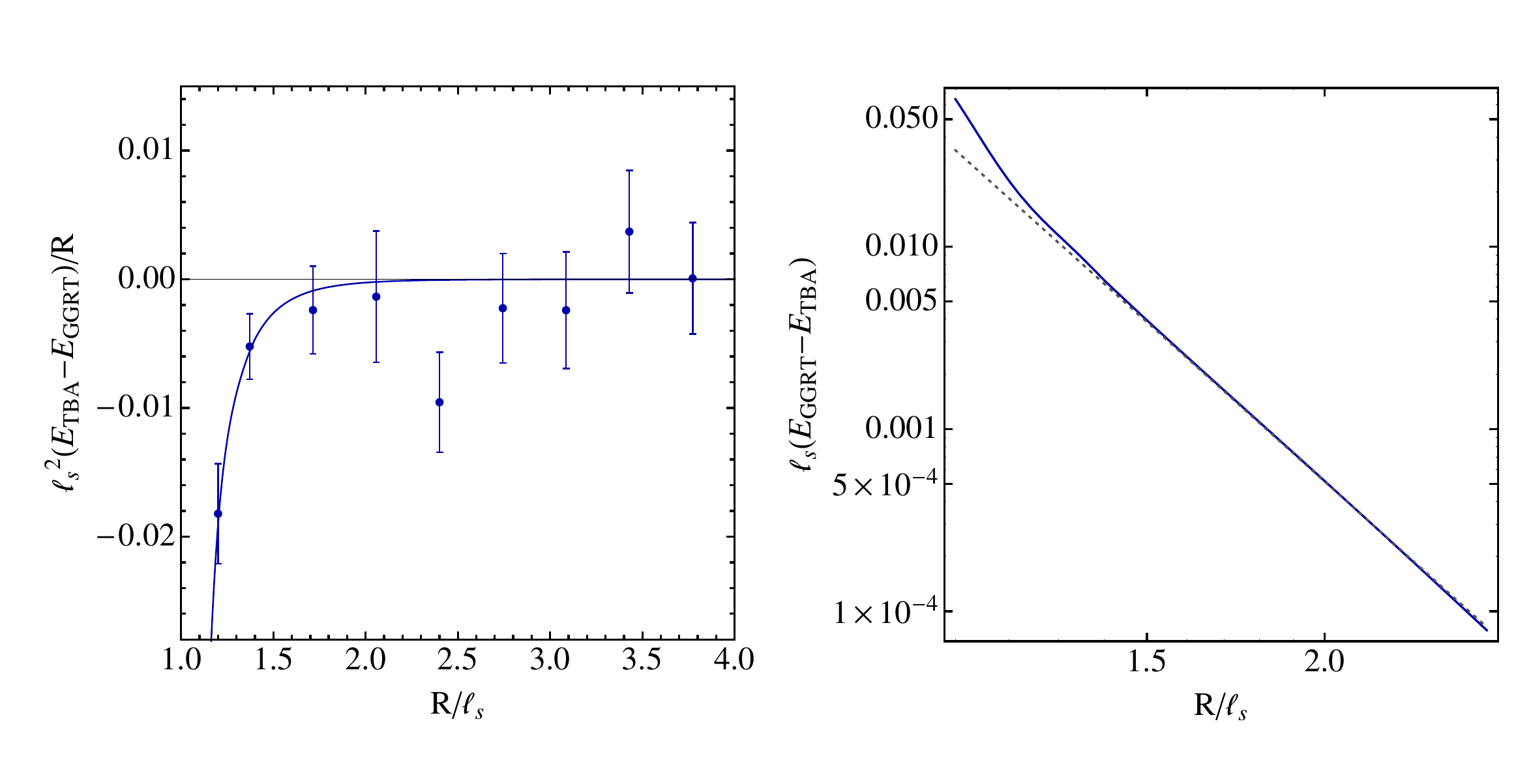} 
 \caption{The left panel shows the ground state energy predicted from the fit of the phase shift to the excited states relative to the GGRT prediction. The right panel shows a log-log plot illustrating that the correction for $R>1.5\ell_s$ behaves as $R^{-7}$ and becomes as steep as $R^{-11}$ for smaller $R$.}
 \label{fig:E3dGSgamma}
 \end{center}
\end{figure}

\be
\label{TBAc4}
c=1+2\frac{p\ell_s^2}{R}-\frac{\pi(D-2)}{12 R^2 c}\ell_s^2\,,
\ee
\be
\label{TBAp4}
pR+ 4\delta(p)-\frac{\pi(D-2)}{12Rc}\ell_s^2p=2 \pi N\,,
\ee
with $N=1$, and 
\be
\label{TBAe4}
\Delta E=4p-\frac{\pi(D-2)}{6 R c}\;.
\ee
Including all data points with $p\ell_s\leq2$, and taking the error bars at face value, we find 
\bea\label{eq:gamma3}
\gamma_3=\frac{0.7\pm 0.1}{(2\pi)^2}\,,
\eea
non-zero at approximately 7 $\sigma$. This correction increases the binding energies and thus lowers the energies of the theory prediction. It also introduces a splitting between two-particle states and four-particle states in agreement with the data simply because the phonons comprising the two-particle states carry larger momenta and will be more strongly bound than the phonons making up the four-particle states. 

Ignoring the contributions to the winding corrections from corrections to the GGRT phase shift has so far worked well. However, there is a subtlety at the order in the expansion we are currently working. The positive coefficient $\gamma_3$ implies a correction to the pseudo-energies with a negative coefficient. As a consequence, the integrals in the TBA equations are no longer convergent. These divergences are not surprising and arise because higher derivative theories typically come with ghosts around the cut-off scale. The perturbative calculation presented in section~\ref{sec:diagTBA} shows that this happens for positive $\gamma_3$. We know, of course, that the full theory does not have ghosts and that there are higher order terms that cure the divergences. Introducing such higher order terms by hand seems unsatisfactory because it would introduce additional arbitrary coefficients. It seems more appealing to interptret the $\ell_s^6s^3$ correction as arising from a heavy resonance that has been integrated out, which suggests the phase shift
\bea\label{eq:res}
e^{2i\delta}=e^{2i\tilde\ell_s^2p_lp_r}\frac{s-2iM\Gamma +M^2}{s+2iM\Gamma +M^2}\frac{s-2iM\Gamma - M^2}{s+2iM\Gamma - M^2}\,,
\eea  
where
\bea
\tilde\ell_s^2=\ell_s^2-\frac{32 \Gamma}{M(M^2+4\Gamma^2)}\,,
\eea
so that the correct phase shift is recovered for $s\ll M^2$.
% The interpretation is simply that the string tension in the UV theory containing the heavy state is lower than the effective tension in the IR theory in which the heavy state has been integrated out, and it is the latter that we measure from the ground state. In other words the heavy state is responsible for some of the tension of our effective string. 
This amplitude~\eqref{eq:res} is not consistent with the non-linearly realized symmetries and should for now simply be thought of as a fitting function that has the desirable property that the integrals in the TBA remain finite and corrections to windings relative to those in the GGRT theory remain small. Fitting to the data, we find 
\bea
M=3.7/\ell_s\qquad\text{and}\qquad\Gamma=1.0/\ell_s\,.
\eea 
Upon expansion in $s$, this leads to a value of $\gamma_3$ in a good agreement with equation~\eqref{eq:gamma3} so that the phase shift is well approximated by our fitting function below the resonance. Above the resonance it does not have the correct behavior to be compatible with the non-linearly realized Lorentz invariance and should thus not be trusted for $p\ell_s>1.85$. 

The resulting predictions for the energy levels of the states involved in the fit are shown in Figure~\ref{fig:E3dgamma}, and we see that the modified phase shift correctly reproduces the larger binding energies and the splitting between two- and four-particle states seen in the data. The momenta for some of the data points for the second excited two-particle state as well as the data points for the third excited two-particle state are so large that our approximations become unreliable.
% The right panel in Figure~\ref{fig:E3d} suggests that there is no need to introduce additional ingredients such as massive states to explain the data and can get by with the inclusion of higher order terms, but this deserves a more careful and systematic study. 
The ground state, however, is rather insensitive to the UV behavior of the phase shift, and it is interesting to compute the correction to the ground state energy that corresponds to our correction to the phase shift. We do this by solving the TBA numerically by iteration. The result is shown in Figure~\ref{fig:E3dGSgamma}. The left panel shows that the ground state data is in good agreement with our prediction. The right panel shows that the leading correction to the GGRT ground state energy is well described by a $R^{-7}$ term down to $R>1.5\ell_s$ and becomes as steep as $R^{-11}$ for the shortest strings studied in~\cite{Athenodorou:2011rx}. 

\subsection{$k$-strings in $D=3$ Yang--Mills}
\label{sec:kstrings}
\begin{figure}[t]
 \begin{center}
 \includegraphics[width=5.8in]{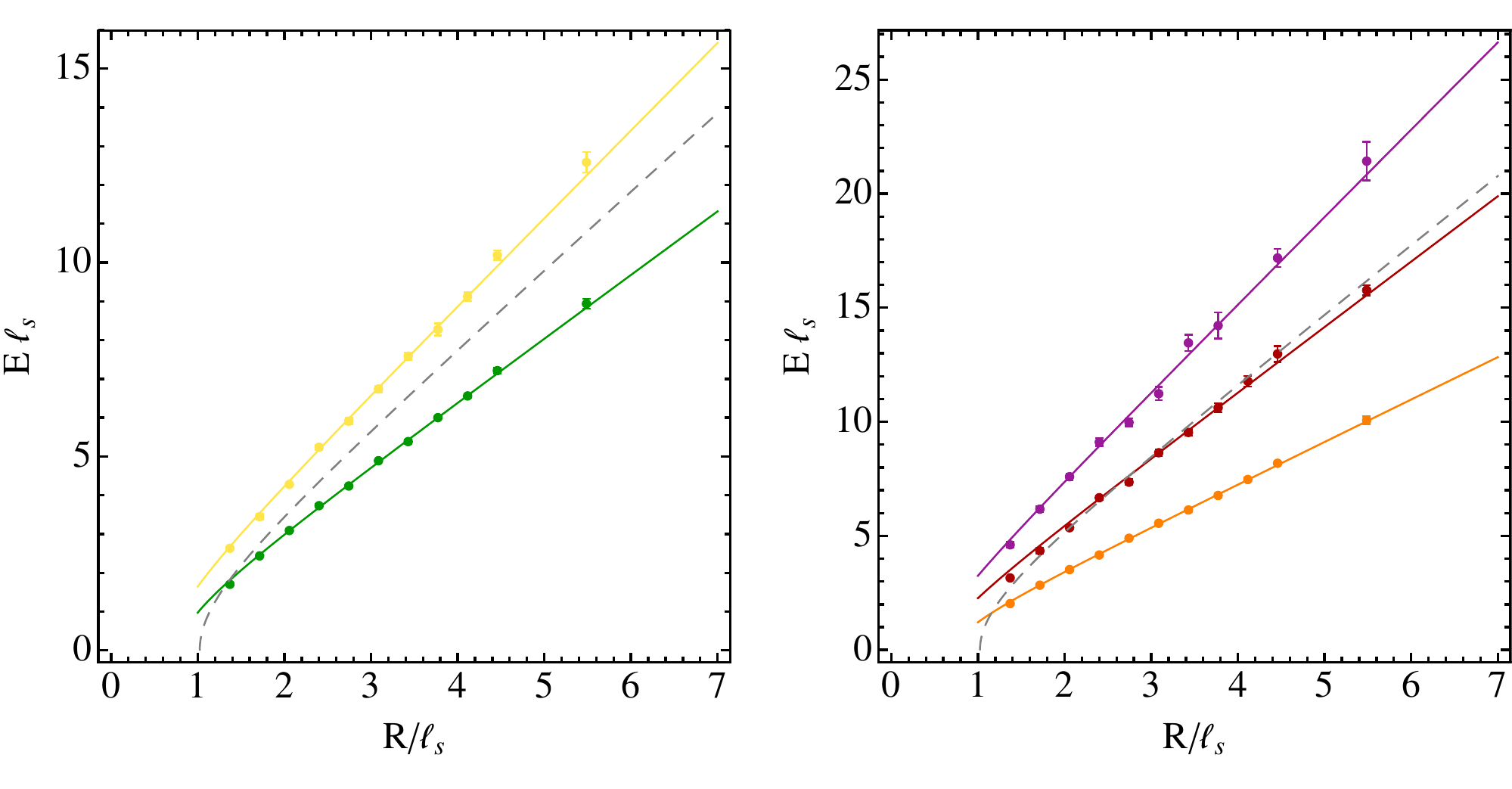} 
 \caption{This plot shows the energies of the ground states for k-strings with $k=2$ (left) and $k=3$ (right). For $k=2$, the data for the anti-symmetric and symmetric representations of SU(6) are shown in green and yellow. For $k=3$, we show the data for the anti-symmetric, mixed, and symmetric representations in orange, red, and purple. The solid lines represent the GGRT prediction with the tension derived from a fit to the data as before. The dashed lines show the energy for the state with the same charge under the center group consisting of non-interacting fundamental strings.}
 \label{fig:GGRTk}
 \end{center}
\end{figure}

\begin{figure}[t]
 \begin{center}
 \includegraphics[trim=1.5cm .1cm 1.5cm .1cm,width=6in]{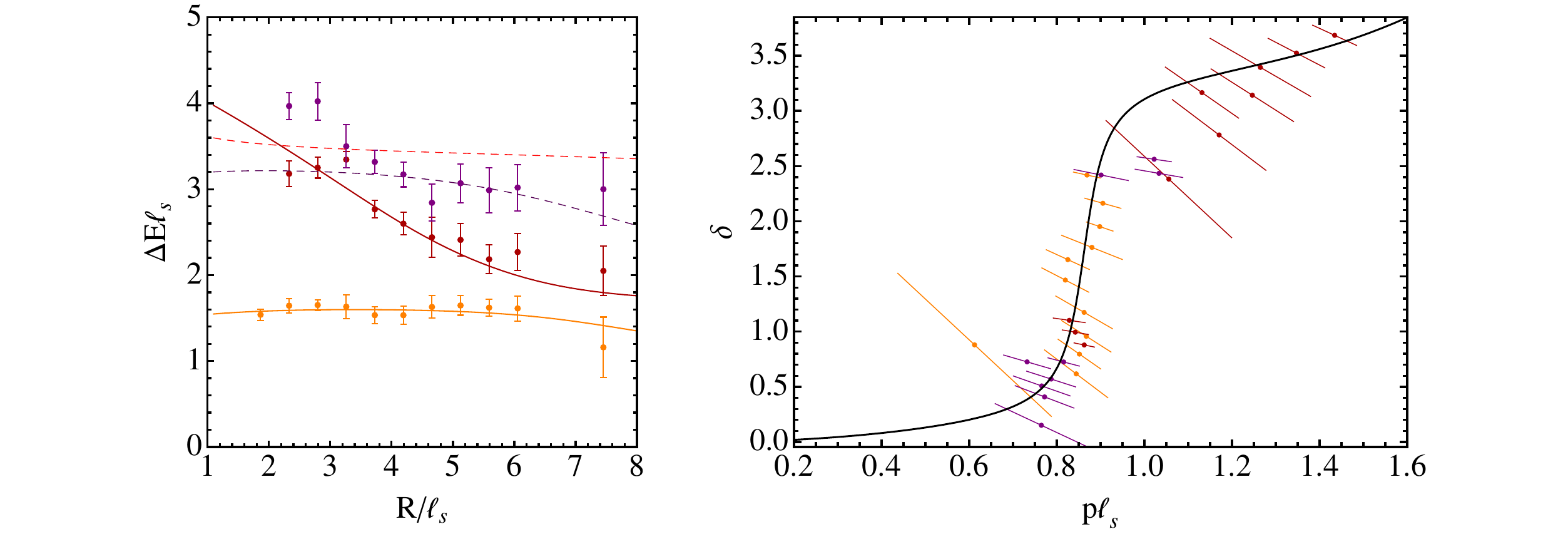} 
 \caption{The left panel shows the energy as a function of string length for the lowest lying excited states for the anti-symmetric representation with $k=3$ with an even number of phonons and zero total momentum. The solid lines are the theory predictions for the 2-particle states, dashed lines represent 4-particle states. The right panel shows the phase shift extracted from the data.}
 \label{fig:k3A}
 \end{center}
\end{figure}

\begin{figure}[t]
 \begin{center}
 \includegraphics[trim=1.5cm .1cm 1.5cm .1cm,width=6in]{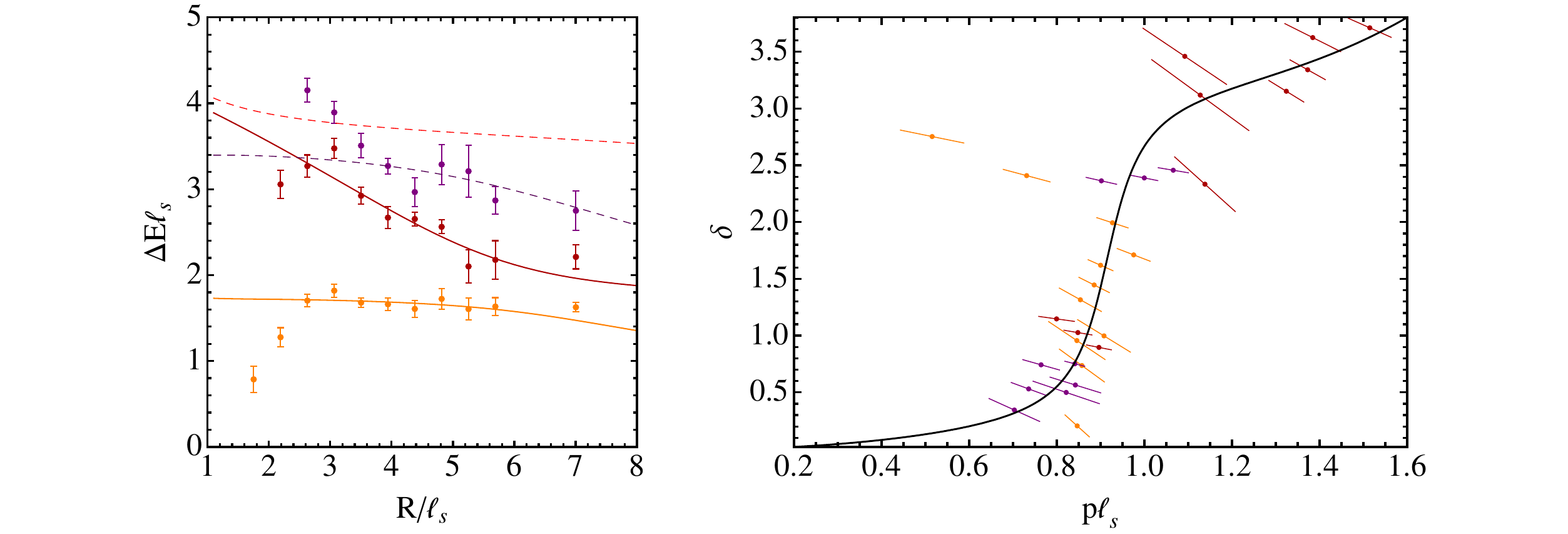} 
 \caption{The left panel shows the energy as a function of string length for the lowest lying excited states for the anti-symmetric representation with $k=2$ with an even number of phonons and zero total momentum. The solid lines are the theory predictions for the 2-particle states, dashed lines represent 4-particle states. The right panel shows the phase shift extracted from the data.}
 \label{fig:k2A}
 \end{center}
\end{figure}

In addition to the data for Wilson loops in the fundamental representation presented in~\cite{Athenodorou:2011rx}, nice data for SU(6) gauge group at $\beta=171$ has recently been presented for bound states of such strings with $k=2$ and $k=3$ units of charge under the center symmetry~\cite{Athenodorou:2013ioa}.  The left panel of Figure~\ref{fig:GGRTk} shows the data for the ground states with $k=2$ in the anti-symmetric and symmetric representation together with the GGRT prediction. The right panel of Figure~\ref{fig:GGRTk} shows the ground state data for the anti-symmetric, mixed, and symmetric representations with $k=3$. Higher representations are related to these by charge conjugation. For comparison we also show the energies corresponding to two and three non-interacting fundamental strings as dashed lines. An analytic calculation of the tension of these objects that is in remarkable agreement with the numerical data can be found in~\cite{Karabali:2007mr}.
The anti-symmetric representations are bound for both $k=2$ and $k=3$. The symmetric representations are unbound for both $k=2$ and $k=3$, while the mixed representation for $k=3$ is at best marginally bound. This motivates us to study the anti-symmetric representations in more detail and leave the others for a future study. To illustrate that our methods also work for $k$-strings, we show the energy levels and phase shift for the states with equal numbers of left- and right-movers and zero total momentum in Figures~\ref{fig:k3A} and~\ref{fig:k2A}, extracting phase shifts for 2- and 4-particle states using equations~\eqref{TBAc}-\eqref{TBAe} and \eqref{TBAc4}-\eqref{TBAe4}. 
The data shows clear evidence for a resonance and the theory predictions are obtained with a resonance with
\bea
m=1.88/\ell_s^{2A}&\text{and}&\Gamma=0.29/\ell_s^{2A}\,,\\
m=1.74/\ell_s^{3A}&\text{and}&\Gamma=0.16/\ell_s^{3A}\,,
\eea
where the superscript denotes the representation of the string. It was perhaps natural to expect the presence of resonances for $k$-strings, given that these can be thought of as bound states of two fundamental flux tubes. It is intriguing that the values for the mass and the width are close when measured in the corresponding string units (and close to the mass and the width of the worldsheet axion in 4D).

These states nicely illustrate that the energy plots can be rather complex because of level crossing even with a very simple phase shift. The solid lines represent the theory predictions for 2-particle states, the dashed lines those for 4-particle states. We clearly see avoided level crossing for the 2-particle states as well as between the 4-particle states. However, the 2- and 4-particle states shown in red and purple cross. In the integrable theory these states have different quantum numbers and do not mix. In QCD integrability is not exact and a certain amount of mixing between 2- and 4-particle states is expected which would lead to avoided level crossing. 

The theory predictions also show that the extraction of these energy levels is very subtle because several energy levels have comparable energies and the correlation function may not be dominated by a single exponential. Also the phase shift extraction from the energy levels in the region of level crossing is not completely straightforward 
due to ambiguities of quantum number assignments. The identification of two- and four-particle states employed here appears to produce the most meaningful results on the phase shift plot, but we cannot exclude at the moment that some of the data points might have been misidentified, especially for  $k=2$ strings. This motivates further 
high precision lattice measurements of these states. Hopefully, techniques presented here might be helpful to guide these measurements.

Notice an interesting feature exhibited by the $k=2$ data---a very pronounced break in the resonance plateau on the energy plot for the lowest (orange) level at $R/\ell_s^{2A}\lesssim 3$.
The corresponding points also show up very far from the theory curve on the corresponding phase shift plot. The natural explanation for the origin of this break is that it occurs when the physical size of the compact dimension becomes comparable to the size of the massive resonant state.  Our phase shift extraction becomes unreliable at these short radii, because the winding corrections due to resonance become large. This interpretation is supported by observing that a very similar break at the same values of $R$
appears also in the lightest glueball energy plot \cite{Athenodorou:2013ioa}, suggesting that the size of the resonance is roughly equal to the size of the lightest glueball.

The $k=3$ data does not exhibit such a break. 
Perhaps only the shortest point in Fig.~\ref{fig:k3A} (with $R/\ell_s^{3A}\approx 2$)  may be considered as an indication for the beginning of the break. This is in agreement with the $k=3$ string being much more strongly bound than the $k=2$ strings. The $k=3$ tension is equal to $\sigma_{3A}\approx 0.6\times 3\sigma_f$, while the $k=2$ tension is
$\sigma_{2A}\approx 0.8 \times 2 \sigma_f$, where $\sigma_f$ is the fundamental flux tube tension.

\section{Future Directions and Conclusions}
\label{sec:last}
We feel that the most important conclusion to be drawn from the current paper is that there is strong motivation for further high precision lattice studies of the properties of flux tubes.
The TBA method provides a solid analytic framework for theoretical interpretation of lattice results for the flux tube lengths which are accessible with the existing computer power. This opens the possibility for a comprehensive description of the world sheet dynamics of the confining strings in the near future, which might be an important step towards understanding the physics of confinement.

The results presented here pose a number of intriguing questions which may be answered with a new data.
Many of them concern the nature of the observed pseudoscalar resonance in the $D=4$ data.
 In particular,  the phase shift plots Figs.~\ref{fig:resonance}, \ref{fig:resKK} show a systematic   disagreement between the theory curve and the data at the momenta above the resonance in the pseudoscalar channel. By itself this disagreement is not very dramatic, given that the corresponding momenta are already quite large. However, an intriguing property of the observed phase shift in the pseudoscalar channel is a pronounced plateau at $\delta\approx\pi$, which corresponds to the absence of scattering.
Together with a systematically better agreement between theory and data in other channels this suggests that some interesting pieces of physics may still be missing.

The experience with $D=3$ $k$-strings data suggest that (at least partially) the plateau may be an artifact resulting from misidentification of the excited state data points. The phase shift plot was constructed assuming these correspond to the two-particle state. It appears very likely that some of these points (in particular, those with $p\ell_s\gtrsim 2$) represent a four-particle state instead. With this interpretation the deduced values of the momenta will be roughly halved, bringing these points in the resonance region and significantly decreasing the tension between theory and data.
Resolving this question will require both more accurate data for excited states in this region and further theoretical work. Indeed, including four-particle states in the analysis is not as straightforward for $D=4$ data as it was in $D=3$ due to a larger number of channels at $D=4$. There is no problem of principle here,  and we are planning to implement this in future work. 

The plateau may be indicative of even more interesting physics. Indeed the axion model (\ref{Lphi}) represents only a minimal effective field theory explaining the observed resonance in the pseudoscalar channel. More complicated scenarios are possible.
In particular, note that as a consequence of two-dimensional kinematics, a two-particle threshold generically appears as a resonant pole. This opens an interesting possibility that the axion may be in fact a threshold bound state of even lighter  massive world sheet excitations. 

\begin{figure}[t]
 \begin{center}
 \includegraphics[width=3in]{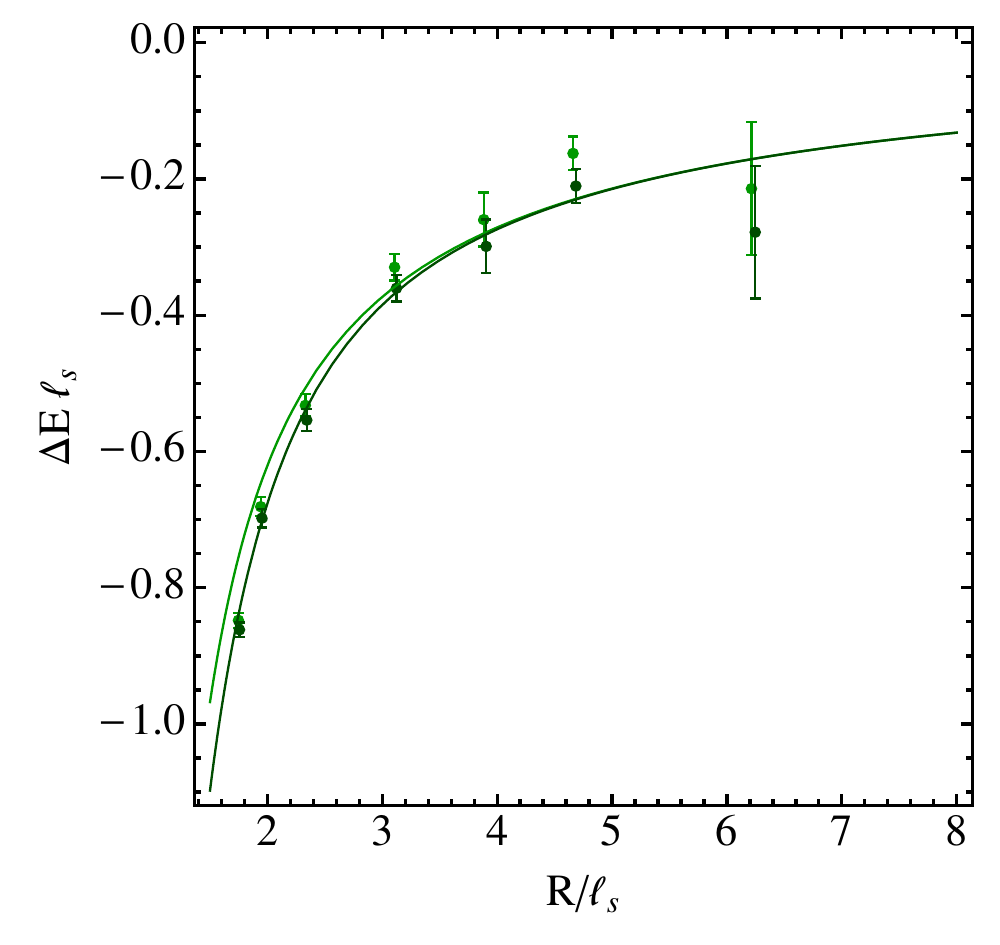} 
 \includegraphics[width=3in]{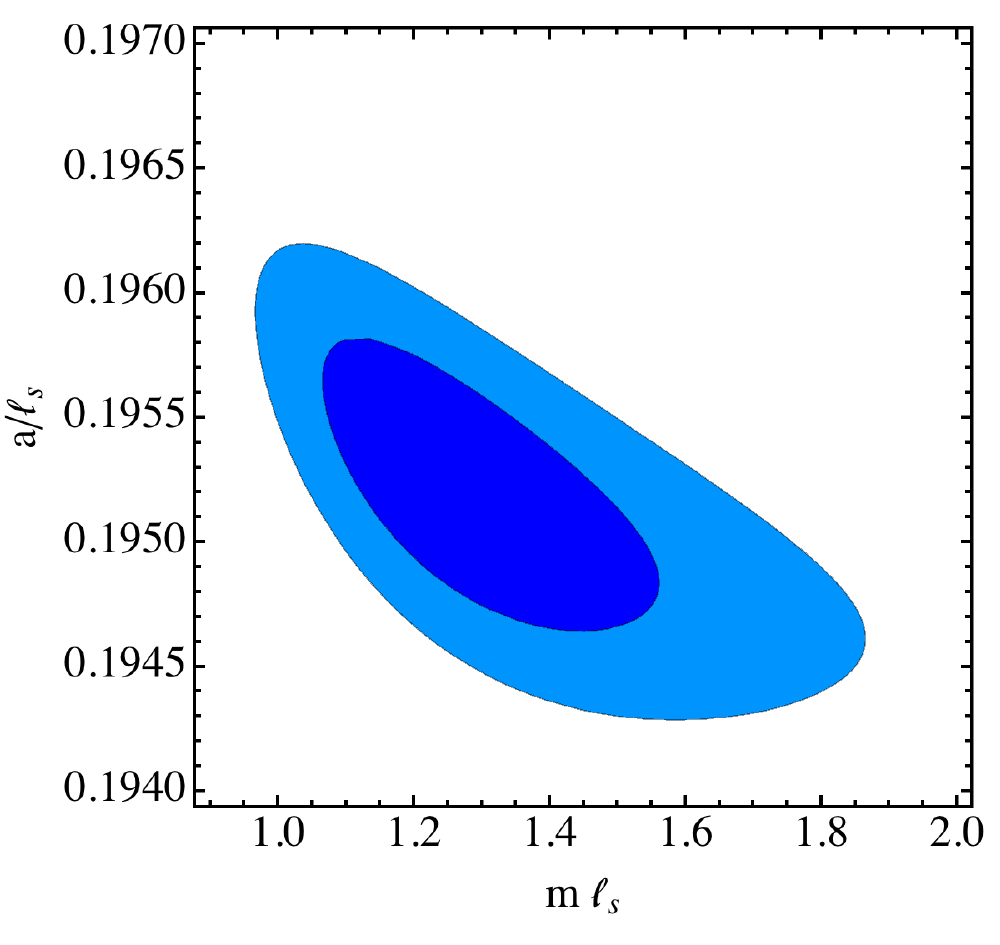} 
 \caption{The left panel shows the ground state energy in the absence and presence of additional massive states in lighter and darker green, respectively. The data points shift becase we simultaneously fit for the mass and the string tension. The right panel shows the one and two sigma contours in the mass-tension plane. The shortest point was not included in the fit, but nevertheless fits well.}
 \label{fig:chi2}
 \end{center}
\end{figure}

Even if this possibility is not realized it is a well-motivated question whether the axion is indeed the lightest massive mode, or there might be lighter massive states missed by the lattice searches. To get an idea of how light these states might be, one can use an available high precision data for the ground state energy. A free particle of a mass $m$ on the world sheet will result in an additional contribution to the ground state Casimir energy of the form
\[
\Delta E(R)=-{m\over \pi}\sum_n \frac1n K_1(mnR)\;.
\]
Given that the lattice data shows no sign of a resonance in the scalar channel we consider the effect of adding a pair of such particles on the ground state energy (having in mind a massive $O(2)$ vector). The result is presented in Fig.~\ref{fig:chi2}. We exclude the data point corresponding to the shortest string from the fit to be conservative. We see that the best fit value for the mass is $m\approx 1.3\ell_s$. Taking the error bars at face value, we find an improvement in the fit corresponding to almost $4\sigma$ (and much larger if the data for the shortest string had been included) in favor of the existence of additional light particles. 

One may achieve a comparable improvement of the fit by adding a $R^{-7}$ correction to the ground state energy, however, the required value for the corresponding coefficient is about a factor of five larger than the typical size of the loop corrections (estimated from the expansion of the GGRT ground state energy).

Of course these considerations do not take into account possible lattice systematics. Most of the improvement in the fit is driven by the data points corresponding to the shortest strings, which might suffer from possible discretization effects and from their proximity to the deconfinement transition.
Nevertheless, this observation provides an additional strong motivation for the systematic search for  ``exotic" light states on the world sheet with quantum numbers, for which the corresponding Nambu--Goto state is expected to be heavy.

The natural candidate operators  for creating new massive states on the world sheet are Polyakov loops with additional local insertions, such as
\be
\label{plaqloop}
W_{\mu\nu}=\Tr\,P\left( F_{\mu\nu}\exp\int_C A\right)\;.
\ee
It is intriguing that the basis of operators used in \cite{Athenodorou:2010cs} includes such an operator with $(\mu\nu)$ indices in the transverse plane ({\it i.e.}, a pseudoscalar), but not with other orientations.
Related to this, to understand the origin of the world sheet axion better it will be interesting to study which operator provides the best overlap for the corresponding state, with (\ref{plaqloop}) providing the most natural candidate.

It would be very interesting to understand the microscopic origin of the world sheet axion, {\it i.e.}, to derive it from the 4D QCD description.
Note in this respect, that a pseudoscalar state with the same mass (in string units) is also present in the available $SU(5)$ data from \cite{Athenodorou:2010cs}, so the axion appears to be present in the large $N$ limit. Unfortunately, however, it appears impossible to use holographic gravitational AdS/QCD models to look for the axion quantitatively. For the gravitational description to be applicable the string length should be short compared to the AdS curvature length. This implies that the mass of the light glueballs (gravitational Kaluza--Klein modes) is parametrically smaller than the confining string tension, which is not the appropriate regime to describe the pure glue theory.
 
A less ambitious goal would be to look for footprints of the world sheet axion in the spectrum of 4D states.
This should be possible,  given that there is no fundamental obstacle for extending the TBA technique to open strings. First steps in this direction have already been made in \cite{Caselle:2013dra} (see also \cite{Hellerman:2013kba}, where the effect of the PS interaction on the open string spectrum was discussed using the conformal gauge approach).
Assuming the world sheet axion survives in the presence of quarks this opens an exciting possibility to see its presence in the physical spectrum of mesons.
In particular, one may expect  mesons with a sufficiently high spin (such that the corresponding flux tube is long enough) to exhibit universal excitations with energy of order the axion mass
$\sim 750$~MeV and of opposite parity, corresponding to an addition of the axion to the confining flux tube. This expectation appears to be supported by the available lattice data for open strings in $SU(3)$ gluodynamics \cite{Juge:2002br}, which shows an anomalous $\Sigma_u^-$ excitation with the energy which matches well the world sheet axion mass, as was previously pointed out by \cite{Andreev:2012mc}.

\section*{Acknowledgments}
We are especially grateful to Mike Teper for many useful discussions and for providing us with the lattice data in electronic form. We thank Ofer Aharony,  Giga Gabadadze, Simeon Hellerman, Zohar Ko\-mar\-godski, Mehrdad Mirbabayi, Joe Polchinski, Massimo Porrati, Matt Roberts, Mi\-sha Shif\-man, Arkady Tsey\-tlin, Gabriele Veneziano and Giovanni Villadoro for useful discussions and feedback. 
This work is supported in part by the NSF under grant numbers PHY-1068438 as well as PHY-1316452. 
RF gratefully acknowledges the Raymond and Beverly Sackler Foundation for their support. 
RF is also supported in part by NSF grants PHY\--1213563 and PHY-0645435.

\bibliographystyle{utphys}
\bibliography{dlrrefs}

\providecommand{\href}[2]{#2}\begingroup\raggedright\begin{thebibliography}{10}

\bibitem{Veneziano:1968yb}
G.~Veneziano, ``{Construction of a crossing - symmetric, Regge behaved
  amplitude for linearly rising trajectories},'' {\em Nuovo Cim.} {\bf A57}
  (1968)
190--197.
%%CITATION = NUCIA,A57,190;%%.

\bibitem{Bissey:2006bz}
F.~Bissey, F.-G. Cao, A.~Kitson, A.~Signal, D.~Leinweber, {\em et al.},
  ``{Gluon flux-tube distribution and linear confinement in baryons},'' {\em
  Phys.Rev.} {\bf D76} (2007) 114512,
\href{http://www.arXiv.org/abs/hep-lat/0606016}{{\tt hep-lat/0606016}}.
%%CITATION = HEP-LAT/0606016;%%.

\bibitem{Maldacena:1997re}
J.~M. Maldacena, ``{The Large N limit of superconformal field theories and
  supergravity},'' {\em Adv.Theor.Math.Phys.} {\bf 2} (1998) 231--252,
\href{http://www.arXiv.org/abs/hep-th/9711200}{{\tt hep-th/9711200}}.
%%CITATION = HEP-TH/9711200;%%.

\bibitem{Polyakov:1998ju}
A.~M. Polyakov, ``{The Wall of the cave},'' {\em Int.J.Mod.Phys.} {\bf A14}
  (1999) 645--658,
\href{http://www.arXiv.org/abs/hep-th/9809057}{{\tt hep-th/9809057}}.
%%CITATION = HEP-TH/9809057;%%.

\bibitem{Athenodorou:2010cs}
A.~Athenodorou, B.~Bringoltz, and M.~Teper, ``{Closed flux tubes and their
  string description in D=3+1 SU(N) gauge theories},'' {\em JHEP} {\bf 1102}
  (2011) 030,
\href{http://www.arXiv.org/abs/1007.4720}{{\tt 1007.4720}}.
%%CITATION = ARXIV:1007.4720;%%.

\bibitem{Athenodorou:2011rx}
A.~Athenodorou, B.~Bringoltz, and M.~Teper, ``{Closed flux tubes and their
  string description in D=2+1 SU(N) gauge theories},'' {\em JHEP} {\bf 1105}
  (2011) 042,
\href{http://www.arXiv.org/abs/1103.5854}{{\tt 1103.5854}}.
%%CITATION = ARXIV:1103.5854;%%.

\bibitem{Athenodorou:2013ioa}
A.~Athenodorou and M.~Teper, ``{Closed flux tubes in higher representations and
  their string description in $D=2+1 SU(N)$ gauge theories},'' {\em JHEP} {\bf
  1306} (2013) 053,
\href{http://www.arXiv.org/abs/1303.5946}{{\tt 1303.5946}}.
%%CITATION = ARXIV:1303.5946;%%.

\bibitem{Luscher:1980ac}
M.~L{\"u}scher, ``{Symmetry Breaking Aspects of the Roughening Transition in
  Gauge Theories},'' {\em Nucl.Phys.} {\bf B180} (1981) 317.

\bibitem{Luscher:2004ib}
M.~L{\"u}scher and P.~Weisz, ``{String excitation energies in SU(N) gauge
  theories beyond the free-string approximation},'' {\em JHEP} {\bf 0407}
  (2004) 014, \href{http://www.arXiv.org/abs/hep-th/0406205}{{\tt
  hep-th/0406205}}.

\bibitem{Aharony:2010db}
O.~Aharony and N.~Klinghoffer, ``{Corrections to Nambu-Goto energy levels from
  the effective string action},'' {\em JHEP} {\bf 1012} (2010) 058,
  \href{http://www.arXiv.org/abs/1008.2648}{{\tt 1008.2648}}.

\bibitem{Goddard:1973qh}
P.~Goddard, J.~Goldstone, C.~Rebbi, and C.~B. Thorn, ``{Quantum dynamics of a
  massless relativistic string},'' {\em Nucl.Phys.} {\bf B56} (1973)
109--135.
%%CITATION = NUPHA,B56,109;%%.

\bibitem{Arvis:1983fp}
J.~Arvis, ``{The Exact Q Anti-Q Potential In Nambu String Theory},'' {\em
  Phys.Lett.} {\bf B127} (1983)
106.
%%CITATION = PHLTA,B127,106;%%.

\bibitem{Dubovsky:2013gi}
S.~Dubovsky, R.~Flauger, and V.~Gorbenko, ``{Evidence for a new particle on the
  worldsheet of the QCD flux tube},'' {\em Phys.Rev.Lett.} {\bf 111} (2013)
  062006,
\href{http://www.arXiv.org/abs/1301.2325}{{\tt 1301.2325}}.
%%CITATION = ARXIV:1301.2325;%%.

\bibitem{Aharony:2013ipa}
O.~Aharony and Z.~Komargodski, ``{The Effective Theory of Long Strings},'' {\em
  JHEP} {\bf 1305} (2013) 118,
\href{http://www.arXiv.org/abs/1302.6257}{{\tt 1302.6257}}.
%%CITATION = ARXIV:1302.6257;%%.

\bibitem{Dubovsky:2012sh}
S.~Dubovsky, R.~Flauger, and V.~Gorbenko, ``{Effective String Theory
  Revisited},'' {\em JHEP} {\bf 1209} (2012) 044,
\href{http://www.arXiv.org/abs/1203.1054}{{\tt 1203.1054}}.
%%CITATION = ARXIV:1203.1054;%%.

\bibitem{Zamolodchikov:1989cf}
A.~Zamolodchikov, ``{Thermodynamic Bethe Ansatz In Relativistic Models. Scaling
  Three State Potts And Lee-Yang Models},'' {\em Nucl.Phys.} {\bf B342} (1990)
695--720.
%%CITATION = NUPHA,B342,695;%%.

\bibitem{Dorey:1996re}
P.~Dorey and R.~Tateo, ``{Excited states by analytic continuation of TBA
  equations},'' {\em Nucl.Phys.} {\bf B482} (1996) 639--659,
\href{http://www.arXiv.org/abs/hep-th/9607167}{{\tt hep-th/9607167}}.
%%CITATION = HEP-TH/9607167;%%.

\bibitem{Luscher:1985dn}
M.~L{\"u}scher, ``{Volume Dependence of the Energy Spectrum in Massive Quantum
  Field Theories. 1. Stable Particle States},'' {\em Commun.Math.Phys.} {\bf
  104} (1986)
177.
%%CITATION = CMPHA,104,177;%%.

\bibitem{Luscher:1986pf}
M.~L{\"u}scher, ``{Volume Dependence of the Energy Spectrum in Massive Quantum
  Field Theories. 2. Scattering States},'' {\em Commun.Math.Phys.} {\bf 105}
  (1986)
153--188.
%%CITATION = CMPHA,105,153;%%.

\bibitem{Lucini:2012gg}
B.~Lucini and M.~Panero, ``{SU(N) gauge theories at large N},'' {\em
  Phys.Rept.} {\bf 526} (2013) 93--163,
\href{http://www.arXiv.org/abs/1210.4997}{{\tt 1210.4997}}.
%%CITATION = ARXIV:1210.4997;%%.

\bibitem{Mezincescu:2010yp}
L.~Mezincescu and P.~K. Townsend, ``{Anyons from Strings},'' {\em
  Phys.Rev.Lett.} {\bf 105} (2010) 191601,
\href{http://www.arXiv.org/abs/1008.2334}{{\tt 1008.2334}}.
%%CITATION = ARXIV:1008.2334;%%.

\bibitem{Dubovsky:2012wk}
S.~Dubovsky, R.~Flauger, and V.~Gorbenko, ``{Solving the Simplest Theory of
  Quantum Gravity},'' {\em JHEP} {\bf 1209} (2012) 133,
\href{http://www.arXiv.org/abs/1205.6805}{{\tt 1205.6805}}.
%%CITATION = ARXIV:1205.6805;%%.

\bibitem{Polchinski:1991ax}
J.~Polchinski and A.~Strominger, ``{Effective string theory},'' {\em
  Phys.Rev.Lett.} {\bf 67} (1991) 1681--1684.

\bibitem{Aharony:2011ga}
O.~Aharony, M.~Field, and N.~Klinghoffer, ``{The effective string spectrum in
  the orthogonal gauge},'' {\em JHEP} {\bf 1204} (2012) 048,
\href{http://www.arXiv.org/abs/1111.5757}{{\tt 1111.5757}}.
%%CITATION = ARXIV:1111.5757;%%.

\bibitem{Hansen:2012tf}
M.~T. Hansen and S.~R. Sharpe, ``{Multiple-channel generalization of
  Lellouch-Luscher formula},'' {\em Phys.Rev.} {\bf D86} (2012) 016007,
\href{http://www.arXiv.org/abs/1204.0826}{{\tt 1204.0826}}.
%%CITATION = ARXIV:1204.0826;%%.

\bibitem{Hansen:2013dla}
M.~T. Hansen and S.~R. Sharpe, ``{Relativistic, model-independent,
  three-particle quantization condition},''
\href{http://www.arXiv.org/abs/1311.4848}{{\tt 1311.4848}}.
%%CITATION = ARXIV:1311.4848;%%.

\bibitem{Teschner:2007ng}
J.~Teschner, ``{On the spectrum of the Sinh-Gordon model in finite volume},''
  {\em Nucl.Phys.} {\bf B799} (2008) 403--429,
\href{http://www.arXiv.org/abs/hep-th/0702214}{{\tt hep-th/0702214}}.
%%CITATION = HEP-TH/0702214;%%.

\bibitem{Janik:2010kd}
R.~A. Janik, ``{Review of AdS/CFT Integrability, Chapter III.5: L\'uscher
  Corrections},'' {\em Lett.Math.Phys.} {\bf 99} (2012) 277--297,
\href{http://www.arXiv.org/abs/1012.3994}{{\tt 1012.3994}}.
%%CITATION = ARXIV:1012.3994;%%.

\bibitem{Cooper:2013kga}
P.~Cooper, ``{St\"uckelberg Fields on the Effective p-brane},'' {\em Phys.Rev.}
  {\bf D88} (2013) 025047,
\href{http://www.arXiv.org/abs/1303.0743}{{\tt 1303.0743}}.
%%CITATION = ARXIV:1303.0743;%%.

\bibitem{Polyakov:1986cs}
A.~M. Polyakov, ``{Fine Structure of Strings},'' {\em Nucl.Phys.} {\bf B268}
  (1986) 406--412.

\bibitem{Mazur:1986nr}
P.~Mazur and V.~Nair, ``{Strings in QCD and Theta vacua},'' {\em Nucl.Phys.}
  {\bf B284} (1987)
146.
%%CITATION = NUPHA,B284,146;%%.

\bibitem{Karabali:2007mr}
D.~Karabali and V.~Nair, ``{The Robustness of the vacuum wave function and
  other matters for Yang-Mills theory},'' {\em Phys.Rev.} {\bf D77} (2008)
  025014,
\href{http://www.arXiv.org/abs/0705.2898}{{\tt 0705.2898}}.
%%CITATION = ARXIV:0705.2898;%%.

\bibitem{Caselle:2013dra}
M.~Caselle, D.~Fioravanti, F.~Gliozzi, and R.~Tateo, ``{Quantisation of the
  effective string with TBA},'' {\em JHEP} {\bf 1307} (2013) 071,
\href{http://www.arXiv.org/abs/1305.1278}{{\tt 1305.1278}}.
%%CITATION = ARXIV:1305.1278;%%.

\bibitem{Hellerman:2013kba}
S.~Hellerman and I.~Swanson, ``{String Theory of the Regge Intercept},''
\href{http://www.arXiv.org/abs/1312.0999}{{\tt 1312.0999}}.
%%CITATION = ARXIV:1312.0999;%%.

\bibitem{Juge:2002br}
K.~J. Juge, J.~Kuti, and C.~Morningstar, ``{Fine structure of the QCD string
  spectrum},'' {\em Phys.Rev.Lett.} {\bf 90} (2003) 161601,
\href{http://www.arXiv.org/abs/hep-lat/0207004}{{\tt hep-lat/0207004}}.
%%CITATION = HEP-LAT/0207004;%%.

\bibitem{Andreev:2012mc}
O.~Andreev, ``{Exotic Hybrid Quark Potentials},'' {\em Phys.Rev.} {\bf D86}
  (2012) 065013,
\href{http://www.arXiv.org/abs/1207.1892}{{\tt 1207.1892}}.
%%CITATION = ARXIV:1207.1892;%%.

\end{thebibliography}\endgroup
\end{document}